\documentclass[trackchanges,twocolumn]{aastex701}

\usepackage{apjfonts}
\DeclareSymbolFont{cmletters}{OML}{cmm}{m}{it}
\DeclareMathSymbol{v}{\mathalpha}{cmletters}{"76}

\usepackage{graphicx}	
\usepackage{amsmath}	
\usepackage{amssymb}	
\usepackage{amsfonts,amsthm,epsfig,float,tabularx}
\usepackage{wrapfig}
\usepackage{soul}
\usepackage{placeins}
\usepackage{xcolor}
\usepackage{hyperref}
\usepackage{comment}
\usepackage{CJK}
\usepackage{multirow}

\newcommand{\appropto}{\mathrel{\vcenter{
  \offinterlineskip\halign{\hfil$##$\cr
    \propto\cr\noalign{\kern2pt}\sim\cr\noalign{\kern-2pt}}}}}

\shorttitle{Super-Eddington Magnetically Arrested Disks}
\shortauthors{Kwan et al.}

\graphicspath{{./}{figures/}}

\begin{document}

\title{Strongly Magnetized Super-Eddington Accretion: How Spin and Accretion Rate Regulate Energy Output and Mass Loss}

\author[orcid=0000-0003-0509-2541]{Tom Man Kwan}
\affiliation{Department of Physics, The University of Hong Kong, Pokfulam Road, Hong Kong}
\affiliation{The Hong Kong Institute for Astronomy and Astrophysics, The University of Hong Kong, Pokfulam Road, Hong Kong}
\email{kwanman2@connect.hku.hk}

\author[orcid=0000-0002-9589-5235]{Lixin Dai}
\affiliation{Department of Physics, The University of Hong Kong, Pokfulam Road, Hong Kong}
\affiliation{The Hong Kong Institute for Astronomy and Astrophysics, The University of Hong Kong, Pokfulam Road, Hong Kong}
\email[show]{lixindai@hku.hk}

\author{Cheuk Kwan Kan}
\affiliation{Department of Physics, The University of Hong Kong, Pokfulam Road, Hong Kong}
\affiliation{The Hong Kong Institute for Astronomy and Astrophysics, The University of Hong Kong, Pokfulam Road, Hong Kong}
\email{zoekan@connect.hku.hk}

\author[orcid=0000-0002-0031-3029]{Zepei Xing}
\affiliation{Center for Interdisciplinary Exploration \& Research in Astrophysics (CIERA), Evanston, IL 60208, USA}
\affiliation{Département d’Astronomie, Université de Genève, Chemin Pegasi 51, CH-1290 Versoix, Switzerland}
\affiliation{Gravitational Wave Science Center (GWSC), Université de Genève, Genève, Switzerland}
\email{Zepei.Xing@unige.ch}

\author[orcid=0000-0003-1474-1523]{Tassos Fragos}
\affiliation{Département d’Astronomie, Université de Genève, Chemin Pegasi 51, CH-1290 Versoix, Switzerland}
\affiliation{Gravitational Wave Science Center (GWSC), Université de Genève, Genève, Switzerland}
\email{Anastasios.Fragkos@unige.ch}

\author[orcid=0000-0002-8183-2970]{Matthew Middleton}
\affiliation{School of Physics \& Astronomy, University of Southampton, Southampton, Southampton SO17 1BJ, UK}
\email{M.J.Middleton@soton.ac.uk}

\author[0009-0004-2113-3096] {Tao Ji} 
\affiliation{CDT Data Intensive Science and Industry, Department of Physics and Astronomy, University College London, Gower Street, London WC1E 6BT, UK}
\affiliation{Mullard Space Science Laboratory, University College London, Holmbury St Mary, Dorking, Surrey RH5 6NT, UK}
\affiliation{Department of Physics, The University of Hong Kong, Pokfulam Road, Hong Kong}
\email{jitao@connect.hku.hk}

\author[orcid=0000-0003-3564-6437]{Feng Yuan}
\affiliation{Center for Astronomy and Astrophysics and Department of Physics, Fudan University, Shanghai 200438, People’s Republic of China}
\email{fyuan@fudan.edu.cn}


\begin{abstract}
Strongly magnetized super-Eddington accretion flows power many important astrophysical systems, but how black hole parameters control their output is unclear. We present 30 general relativistic radiation magnetohydrodynamics simulations of super-Eddington magnetically arrested disks onto stellar-mass black holes, varying mass ($M_{\rm BH}= 5, 15, 30\,M_{\odot}$), spin ($a=0.0,0.9$), and accretion rate ($\dot{M}_{\rm acc} \approx 1-2000\,\dot{M}_{\rm Edd}$). We find that black hole spin and accretion rate jointly regulate wind loss rates and energy output efficiencies, while black hole mass has no effect over the mass range studied here. The BH accretes only $7-30\%$ of the mass supplied to the accretion flow, while the rest is expelled in winds. This accretion fraction decreases with mass supply rate and is lower for high-spin systems.  
Both spin states produce strong magnetically driven outflows. For $a = 0$, the wind kinetic, radiative, and electromagnetic efficiencies are modest and show little variation across the full simulated range of accretion rates. For $a = 0.9$, both wind power and jet power increase super-linearly with $\dot{M}_{\rm acc}$, with the jet power saturating beyond $\dot{M}_{\rm acc} \sim 100\,\dot{M}_{\rm Edd}$. Radiation is strongly beamed along the funnel, with inverse beaming factors exceeding $100$ for high-spin, high-$\dot{m}$ models viewed face-on. Our results establish that rapid BH spin boosts energy-extraction efficiency, while high accretion rate amplifies total power. We provide scaling relations for luminosities, jet power, accretion ratio, and beaming, offering a framework for interpreting observations of ULXs and other super-Eddington systems.

\end{abstract}



\section{Introduction} 
\label{sec:intro}

Super-Eddington accretion onto black holes (BHs) powers a diverse range of luminous accreting systems and transients, including ultraluminous X-ray sources \citep[ULXs, see review by][]{Kaaret.2017,King.2023}, tidal disruption events \citep{Evans.1989,Rees.1988,Dai.2021}, narrow-line Seyfert 1 galaxies \citep{Wang.2013,Du.2014}, and microquasars (e.g. SS433, \citealt{Margon.1984,Begelman.2006}; GRS1915+105, \citealt{Fender.2004b}). Their powerful outflows carry mass and energy away from the accretion flow, shaping both the emerging radiation and the feedback on the surrounding environment. Yet how BH mass, spin, and accretion rate regulate these outflows and radiative output remains poorly constrained.

Early models, such as the slim disk \citep{Abramowicz.1988, Sadowski.2009}, predicted that radiative luminosity increases logarithmically, staying at a few times the Eddington luminosity even at extremely high mass accretion rates. Observations of ULXs, however, often exceed this limit \citep{Gladstone.2009}. The luminosity discrepancy is proposed to be partly resolved by beaming models \citep{King.2001, King.2008}, where most radiation escapes through a low-density funnel along the rotation axis, making sources appear super-Eddington when viewed face-on. These early models also did not account for the powerful winds launched from super-Eddington disks. 
Such winds can reach ultrafast velocities while carrying substantial kinetic and radiative power \citep{King.2003,Tombesi.2010, King.2015,Yang.2023}, and provide mechanical feedback capable of inflating observed ULX nebulae \citep{Pakull.2002, Feng.2008, Pakull.2008, Pinto.2016}.

Magnetic fields also play a crucial role in these flows. When the magnetic flux threading the inner disk becomes sufficiently strong, the flow enters the magnetically arrested disk (MAD) state \citep{Narayan.2003}, in which magnetic pressure partially impedes the accretion flow. 
This contrasts with the standard and normal evolution (SANE) state, in which the accumulated magnetic field does not arrest the accreting gas near the BH.
In the MAD state, a rapidly spinning black hole can launch a relativistic jet via the Blandford-Znajek mechanism \citep{Blandford.1977}, with jet power proportional to the square of the spin and the magnetic flux \citep{Tchekhovskoy.2010}. 
Despite the potential importance of magnetic fields, super-Eddington accretion in the MAD regime remains largely underexplored, especially in a systematic way across different black hole parameters.

Previous numerical simulations of super-Eddington accretion (e.g., \citealt{Ohsuga.2005, Jiang.2014, Jiang.2019, Sadowski.2014, Sadowski.2015b, McKinney.2014, Dai18}, also see review by \citealt{Jiang.2024}) have significantly advanced our understanding, revealing that  geometrically thick disks, radiation-driven winds, and, in some cases, relativistic jets, can be formed. However, these studies used different initial conditions, opacity prescriptions, radiation closures, and spanned a range of dimensionalities from two-dimensional axisymmetric configurations to fully three-dimensional simulations. The simulation domains vary from a few tens of to $10^4$ gravitational radii, potentially affecting the treatment of large-scale outflow dynamics and the boundary conditions imposed on the flow. The magnetic field configurations also differ across simulations, with some adopting strong coherent fields that lead to MADs, while others seed weak, loop fields that result in SANE disks. Moreover, each study covered only a limited range of parameters, which did not allow them to systematically explore how key physical parameters such as black hole mass, spin, and accretion rate collectively influence the flow structure, outflow energetics, and radiation.

\begin{figure}
	\includegraphics[width=\columnwidth]{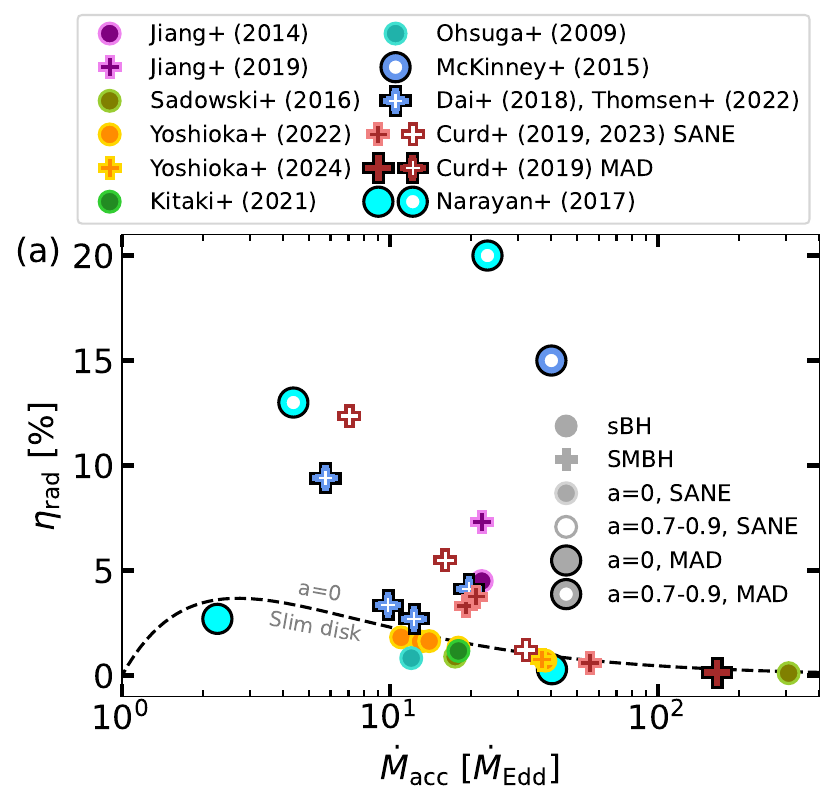}
    \includegraphics[width=\columnwidth]{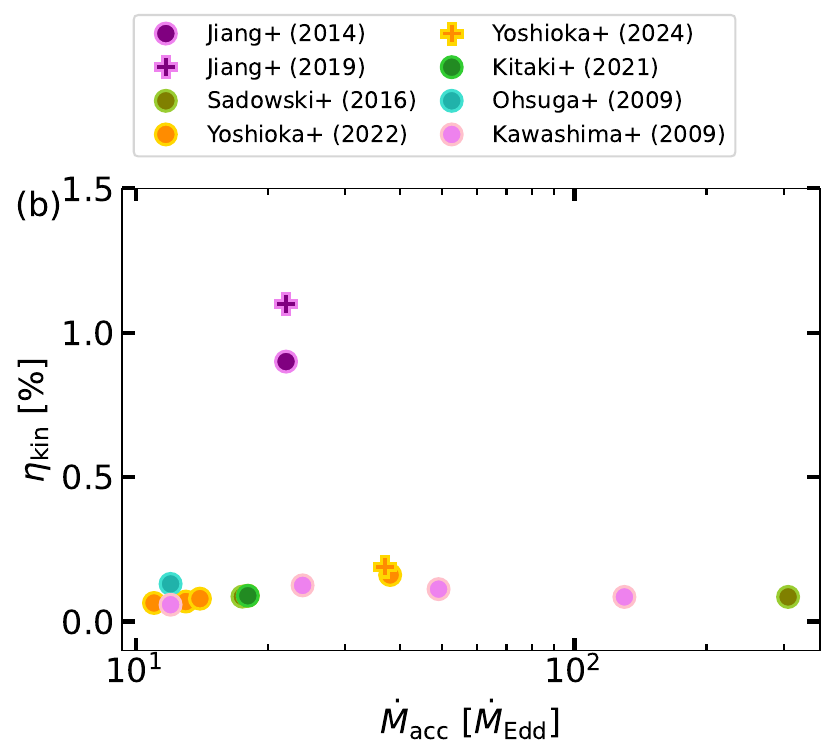}
    \vspace*{-0.4cm}
    \caption{ 
    Summary of previously published results from simulations of super-Eddington BH accretion. Panel (a) shows the radiative efficiency $\eta_{\rm rad}$ and panel (b) shows the kinetic efficiency $\eta_{\rm kin}$ as functions of the BH accretion rate. The black dashed curve in panel (a) shows the slim disk prediction for a non-spinning BH. Colors identify the source simulation; circles and crosses denote stellar-mass and supermassive BHs, respectively; filled and open symbols indicate non-spinning and rapidly spinning BHs, respectively; and black edges mark MAD simulations. The accretion rates have been converted to the Eddington units defined in this work (Equation~\ref{eq:Medd}). To our knowledge, no previous MAD simulation has reported a pure kinetic luminosity.}
    \label{fig:literature}
\end{figure}

Figure~\ref{fig:literature} compiles representative published values of the radiative efficiency $\eta_{\rm rad}$ (Equation~\ref{eq:eta_rad}) and kinetic efficiency $\eta_{\rm kin}$ (Equation~\ref{eq:eta_kin}). The figure shows the broad scatter in these efficiencies across different simulation sets. In some cases, such scatter exists even when the BH spin and accretion rates are similar between two simulations, demonstrating the sensitivity of the reported efficiencies to differences in numerical setup and physical prescriptions.
Several simulations report radiative efficiencies that substantially exceed the standard slim-disk predictions, particularly in the MAD cases, highlighting the role of strong magnetic fields in enhancing the energy output. However, we are not aware of any previous MAD simulation that reports the kinetic efficiency of the wind. In summary, the dependence of these energy efficiencies on key parameters thus remains poorly understood.

Motivated by this, we carry out a large suite of three-dimensional general relativistic radiation magnetohydrodynamics (GRRMHD) simulations of super-Eddington MADs around stellar-mass black holes, systematically varying BH mass ($M_{\rm BH}=5$, $15$, and $30\,M_\odot$), spin ($a=0$ and $0.9$), and mass accretion rate ($\dot{M}_{\rm acc} \approx 1-2000\,\dot{M}_{\rm Edd}$, where $\dot{M}_{\rm Edd}$ is the Eddington rate, Equation \ref{eq:Medd}). Our goal is to determine how these parameters collectively influence the disk structure, wind and jet formation, and radiative and kinetic output, and to provide quantitative scaling relations that can be compared with observations of ULXs and other super-Eddington systems.

The paper is organized as follows. Section \ref{sec:methods} describes our numerical methods. Section \ref{sec:results} presents the results, including flow structure,  outflow energy output, mass budget, and radiation properties. Section \ref{sec:discussion} discusses implications for ULXs and their nebulae, compares with previous simulations, and outlines limitations. Section \ref{sec:summary} summarizes our findings.

\section{Numerical Methods}
\label{sec:methods}

\subsection{GRRMHD Code}
\label{sec:setup}

We carry out 3D simulations of super-Eddington disks using the GRRMHD code {\tt HARMRAD} with M1 closure \citep{Sadowski.2013,McKinney.2014}.
{\tt HARMRAD} solves the coupled conservation equations for rest-mass density, gas+electromagnetic stress-energy, and radiation stress-energy. Gas and radiation interact via the radiative four-force. 
The code has been extensively tested and used to simulate a number of super-Eddington accretion disks involving dynamically important magnetic fields \citep{McKinney.2015, Dai18, Thomsen.2022}.

We employ modified spherical polar coordinates centered on the BH \citep{McKinney.2012,McKinney.2015}, with radial grid spanning from $R_{\rm in}=0.85R_H$ to $R_{\rm out}=8500\,r_g$ (where $r_g = GM_{\rm BH}/c^2$ is the gravitational radius and $R_H$ is the horizon radius). The grid uses static mesh refinement, with cell size increasing exponentially out to a break radius $R_{\rm break}=500\,r_g$ and hyper-exponentially beyond in the r direction. The $\theta$-grid spans $0$ to $\pi$ with finer resolution in the jet and disk regions, while the $\phi$-grid spans $0$ to $2\pi$ with periodic boundary conditions. 

All simulations produce snapshots every $\sim 4 t_g$ and are evolved for at least $t = 30,000\,t_g$ (where $t_g = r_g/c$), reaching quasi-steady inflow out to $r\gtrsim100\,r_g$. 
Grid resolutions $N_r\times N_\theta\times N_\phi$ and simulation durations $T_{\rm max}$ for each model are listed in Table~\ref{tab:models}.
Numerical floors and ceilings are applied to prevent vacuum violations in the jet region. 
More details of the code are provided in Appendix~\ref{ap:setup}.
Furthermore, we refer the reader to Appendix~\ref{ap:mri} for calculations of the magnetorotational instability (MRI) quality factor, where we show that the MRI is well resolved in the quasi-steady inflow in all models.

The gas (plasma) is assumed to be fully ionized and to have solar chemical abundances, which have the mass fractions of H, He, and ``metals'', $X = 0.7, Y = 0.28, Z = 0.02$, respectively. The electron scattering opacity $\kappa_{\rm es}$ and Planck-mean absorption opacity $\kappa_{\rm abs}$, including bound-free, free-free, molecular, H$^-$ and ``Chianti'' opacities, are used as described in \citet{McKinney.2015} (see Appendix~\ref{ap:setup} for details). The total opacity is given by $\kappa_{\rm tot}=\kappa_{\rm abs}+\kappa_{\rm es}$. Thermal Comptonization is also included.

\subsection{Initial conditions}
\label{sec:initial_disks}

We initialize a geometrically thick, Keplerian disk following \citet{Dai18}, except that we do not truncate it at an outer radius.
The rest-mass density of the gas is described by
\begin{equation}
    \rho(r,z) = \rho_0 r^{-1.3}  e^{-z^2/(2H^2)},
\end{equation}
where $H\equiv 0.3R=0.3r\sin\theta$ is the initial disk scale height, $z\equiv r\cos\theta$, and $\rho_0$ is the initial reference density. Varying $\rho_0$ produces different BH mass accretion rates across our simulation suite.

The total pressure of the gas and  radiation fluid $p_{\rm tot}$ is set by local equilibrium assuming an effective adiabatic index $\gamma_{\rm tot}=4/3$ as
\begin{equation}
    p_{\rm tot} = \frac{\rho (H/R)^2 v_K^2}{\gamma_{\rm tot}},
\end{equation}
where $v_K \approx 1/(r^{1.5}+a)$ is the Keplerian velocity. The pressure is randomly perturbed by $10\%$ to seed the magnetorotational instability (MRI) and then partitioned between gas and radiation assuming local thermal equilibrium at a single temperature $T$:
\begin{equation}
    p_{\rm tot} = p_g + p_{\rm rad} = \frac{k_B\rho T}{\mu m_p} + \frac{1}{3}a_{\rm rad}T^4,
\end{equation}
where $p_{g}$ is the gas pressure, $p_{\rm rad}$ is the radiation pressure, $a_{\rm rad}$ is the radiation constant, $k_B$ is the Boltzmann's constant, $m_p$ is the proton mass and $\mu=(2X+0.75Y+0.5Z)^{-1}$ is the mean molecular weight.
The gas internal energy density follows as $u_g = p_g/(\gamma_g-1)$ with $\gamma_g=5/3$. Initial radiation fluxes are then set via flux-limited diffusion as described in Section 6.3 in \cite{McKinney.2014}.

A large-scale poloidal magnetic field aligned with the BH spin axis is initialized following the vector potential prescription of \citet{McKinney.2015}. The field is normalized so that the initial plasma beta $\beta \equiv (p_g+p_{\rm rad})/p_b \approx 30$ along the disk midplane. This initially weak field rapidly evolves into a MAD state as accretion proceeds.

\subsection{Eddington luminosity and accretion rate}
\label{sec:edd}

The Eddington luminosity is defined as
\begin{equation}
    L_{\rm Edd} \equiv \frac{4\pi GM_{\rm BH}c}{\kappa_{\rm es}} \approx 1.3\times 10^{38} \frac{M_{\rm BH}}{M_{\odot}}\ \rm erg\ s^{-1},
\end{equation}
where $\kappa_{\rm es}$ is the electron scattering opacity, $c$ is the speed of light, $G$ is the gravitational constant, and $M_{\rm BH}$ is the black hole mass in units of solar mass $M_{\odot}$.

The corresponding Eddington accretion rate is
\begin{equation}
\label{eq:Medd}
\dot{M}_{\rm Edd} \equiv \frac{L_{\rm Edd}}{\eta_{\rm acc} c^2},
\end{equation}
where $\eta_{\rm acc}$ is the radiative efficiency of the accretion flow. Throughout this work, we adopt a fixed radiative efficiency $\eta_{\rm acc}=0.1$ to enable consistent comparison of absolute mass fluxes across our simulation suite.

We also define the dimensionless Eddington ratio parameter
\begin{equation}
\dot{m} \equiv \dot{M}_{\rm acc}/\dot{M}_{\rm Edd}.
\end{equation}

\subsection{Key diagnostics}
\label{sec:diag}
Below we define the key diagnostics used throughout this paper.

\subsubsection{Mass fluxes}

The gas rest-mass flux flowing inward through a sphere of radius $r$ is
\begin{equation}
	\dot{M} (r)= -\oint \rho u^r \rm dA_{\rm\theta\phi}, \label{eq:mdot}
\end{equation} 
where d$A_{\theta\phi}=\sqrt{-g} {\rm d} \theta {\rm d}\phi$.

Evaluating this flux at the BH event horizon gives the BH accretion rate $\dot{M}_{\rm acc}$. In practice, we measure $\dot{M}_{\rm acc}$ at $r=4r_g$ to avoid contamination from numerical density floors very close to the horizon, since in a quasi-steady state the true accretion rate should remain constant near the horizon.
The wind mass loss rate is defined similarly:
\begin{equation}
    \dot{M}_{\rm wind} = \int_{\rm wind} \rho u^r (v^r>0) \, \mathrm{d}A_{\theta\phi},
\end{equation}
where $v^r \equiv \sqrt{g_{\rm rr}}u^r/u^t$ is the lab frame radial velocity.
We define the total gas mass supply rate as $\dot{M}_t \equiv \dot{M}_{\rm acc} + \dot{M}_{\rm wind}$.

While there is no perfect definition for a ``true wind'' that never falls back onto the BH, we include only the thermally unbound component ($-hu_t>1$), where $h = 1 + \gamma u_g/\rho$ is the gas specific enthalpy. This criterion selects gas that is gravitationally unbound. We refer the reader to Section~\ref{sec:structure} for more details on the flow structure and justification of this choice.

\subsubsection{Energy fluxes}

The total energy flux flowing outwards through a sphere of radius $r$ is 
\begin{equation}
    \dot{E} (r) = -\oint T^r_t \rm dA_{\rm\theta\phi} , \label{eq:edot}
\end{equation} 
where $T^r_t$ is the stress-energy tensor of the ideal MHD fluid.

For a selected region $\mathcal{R}$ of the accretion flow, the electromagnetic (EM) luminosity is
\begin{equation}
    L_{\rm{} EM} (r) = -\int_{\mathcal{R}} {T^{\rm EM}}^r_t {\rm dA_{\theta\phi}}, \label{eq:LEM}
\end{equation}
where ${T^{\rm EM}}^{\mu}_{\nu} = b^2 u^{\mu}u_{\nu} + p_b \delta^{\mu}_{\nu}-b^{\mu}b_{\nu}$ is the EM part of the MHD stress-energy tensor. Here, $p_b=b^{\mu}b_{\mu}/2=b^2/2$ is the magnetic pressure, $b^{\mu}$ and $b_{\nu}$ are the contravariant and covariant magnetic field 4-vectors in the fluid frame, and $\delta^{\mu}_{\nu}$ is the Kronecker delta function. The luminosity measures the outward transport of EM energy in the magnetized outflow. It includes both the advection of magnetic enthalpy by the moving plasma and the transport of energy by magnetic stresses (Poynting flux).

The kinetic luminosity carried by the gas is
\begin{equation}
    L_{\rm kin}(r) = -\int_{\mathcal{R}} (u_t + 1)\rho u^r \, \mathrm{d}A_{\theta\phi},
    \label{eq:Lkin}
\end{equation}
which measures the free kinetic energy flux available at infinity.

The bolometric radiative luminosity is
\begin{equation}
    L_{\rm rad}(r) = -\int_{\mathcal{R}} R^r_t\,  \mathrm{d}A_{\theta\phi},
    \label{eq:Lrad}
\end{equation}
$R^{\mu}_{\nu}$ is the stress-energy tensor of radiation. This quantity measures the outward energy flux carried by the radiation field.

The kinetic, radiative and EM energy efficiencies are defined as:
\begin{equation}
    \eta_{\rm kin} (r) \equiv L_{\rm{}kin} (r)/ \langle \dot{M}_{\rm acc} c^2\rangle_t,
    \label{eq:eta_kin}
\end{equation}
\begin{equation}
    \eta_{\rm rad} (r) \equiv L_{\rm{}rad} (r)/ \langle \dot{M}_{\rm acc} c^2\rangle_t , 
    \label{eq:eta_rad}
\end{equation}
\begin{equation}
    \eta_{\rm EM} (r) \equiv L_{\rm{}EM} (r)/ \langle \dot{M}_{\rm acc} c^2\rangle_t . 
    \label{eq:eta_EM}
\end{equation}
Here $\langle Q \rangle_t$ denotes the time-average of quantity $Q$ over the final period of $t\sim 10000\, t_g$ of each simulation, where the inner disk has reached a quasi-steady state.

The total power measures the net outward energy flux after the rest-mass energy has been subtracted. It is the sum of the radiative, kinetic, and EM luminosities, $L_{\rm total} =L_{\rm EM}+L_{\rm kin}+L_{\rm rad}$. This quantity represents the total energy budget available at infinity. 
For the thermally unbound wind, we measure $L_{\rm wind} = L_{\rm total}(v^r>0,\,-hu_t>1)$. The corresponding total wind efficiency is 
\begin{equation}
    \eta_w (r) \equiv L_{\rm{}wind} (r)/ \langle \dot{M}_{\rm acc} c^2 \rangle_t . 
    \label{eq:eta_wind}
\end{equation}

\subsubsection{Magnetic flux}

The dimensionless magnetic flux through the horizon (``MADness parameter'') is defined as:
\begin{equation}
    \phi_H \equiv \frac{\Phi_H}{\sqrt{\dot{M}_{\rm acc} r_g^2 c}}, 
    \label{eq:mad}
\end{equation}
where $\Phi_H = \frac{1}{2}\oint |B^r| \, \mathrm{d}A_{\theta\phi}$ is the total magnetic flux through the BH event horizon.

The disk enters the MAD state when $\phi_H$ saturates at $\sim 30-50$, depending on the BH spin \citep{Tchekhovskoy.2011, Tchekhovskoy.2012a, Narayan.2022}.

\subsubsection{Jet power}

We can evaluate the jet total power, which is the summation of the jet's MHD energy flux and radiation flux,
\begin{equation}
    P_{\rm{}jet} (r) = -\int_{\rm jet} (T^r_t+R^r_t) (\sigma>1) \ \rm dA_{\rm\theta\phi} , \label{eq:edot_jet}
\end{equation}
within the region of strong magnetization $\sigma \equiv b^2/\rho$. We define the jet efficiency as
\begin{equation}
	\eta_{\rm jet} (r) \equiv P_{\rm{}jet}(r)/ \langle \dot{M}_{\rm acc} c^2\rangle_t .
	\label{eq:eta_jet}
\end{equation}

\subsubsection{Photospheres}
\label{sec:photosphere-def}

We first consider the scattering photosphere, defined by the electron scattering optical depth $\tau_{\rm es}=1$. The electron scattering opacity is $\kappa_{\rm es} \approx 0.2(1+X)\,\rm cm^2\,g^{-1}$, where $X=0.7$ for solar abundance. Because the scattering photosphere can extend beyond our outer boundary at high inclinations due to high gas densities, we compute $\tau_{\rm es}$ along the angular direction at fixed radius:
\begin{equation}
    \tau_{\rm es}(\theta) = \int_{0}^{\theta} \rho \kappa_{\rm es} \, \mathrm{d}l,
\end{equation}
where $\mathrm{d}l = f_\gamma r\mathrm{d}\theta$ and $f_\gamma \approx u^t$ accounts for relativistic effects. We perform the integration angularly to avoid an artificial gas density enhancement along the pole caused by numerical contamination from the jet, which would otherwise skew the location of the scattering photosphere in the low inclinations in a radial calculation. 

The effective photosphere is the surface beyond which thermal radiation can escape without being absorbed. It is defined by $\tau_{\rm eff}=1$ using the effective opacity $\kappa_{\rm eff} = \sqrt{3\kappa_{\rm abs}(\kappa_{\rm abs}+\kappa_{\rm es})}$. We compute the effective optical depth by integrating radially inward from the outer boundary:
\begin{equation}
     \tau_{\rm eff}(r) = \int_{R_{\rm out}}^{r} \rho \kappa_{\rm eff} \, \mathrm{d}l,
\end{equation}
where $\mathrm{d}l = -f_\gamma \mathrm{d}r$ with $f_\gamma \approx u^t(1 - v/c)$. The effective photosphere lies inside the scattering photosphere. We can calculate the effective opacity that includes both free-free and bound-free processes, $\kappa_{\rm abs}=\kappa_{\rm ff}+\kappa_{\rm bf}$, which are the dominant absorption processes in super-Eddington flows. 

However, the bound-free process is generally not important for thermalization in the hot, highly ionized super-Eddington outflows in low inclinations studied here, where they carry out the most energy from the accretion flows. Therefore, for the purpose of this work, we define the effective photosphere where the effective opacity only include free-free absorption, that $\kappa_{\rm abs} = \kappa_{\rm ff}$.

\subsubsection{Luminosity and beaming}
\label{sec:beaming}

The bolometric radiative luminosity escaping out of the accretion flow is
\begin{equation}
    L_{\rm rad}(r) = -\int_{\mathcal{R}} R^r_t (\tau_{\rm eff}<1) \, \mathrm{d}A_{\theta\phi},
\label{eq:thin_rad_L}
\end{equation}
where we only include regions outside the effective photosphere ($\tau_{\rm eff}<1$), assuming scattered photons in such regions will eventually escape.

To quantify the anisotropic escape of radiation through the polar funnel, we define the beaming factor $b$ and the isotropic equivalent radiative luminosity $L_{\rm rad,iso}$ as
\begin{equation}
    b(r,\theta) = \frac{L_{\rm rad}(r)}{L_{\rm rad,iso}(r,\theta)}, \qquad
    L_{\rm rad,iso}(r,\theta) = -4\pi r^2 R^r_t.
\end{equation}
The inverse beaming factor $b^{-1}$ gives the factor by which the observed isotropic-equivalent luminosity exceeds the true luminosity. $b^{-1}\gg1$ indicates strong beaming.

\begin{table*}
	\centering
	\caption{Model parameters: Spin $a$, black hole mass $M_{\rm BH}$, average dimensionless magnetization parameter $\phi_H$, BH accretion rate $\dot{M}_{\rm acc}$, wind-loss rate  $\dot{M}_{\rm wind}$ measured for the total outflow at $r=100 r_g$ and for the thermally unbound wind at $r=2000 r_g$, numerical resolution $N_r \times N_{\theta} \times N_{\phi}$ and simulation duration $T_{\rm max}$ }
	\label{tab:models}
	\begin{tabular}{lcccccccr} 
		\hline
        \hline
        
        \multicolumn{1}{c}{\multirow{2}{*}{Model}} &
        \multirow{2}{*}{$a$} &
        \multirow{2}{*}{$M_{\rm BH}$ [$M_{\odot}$]} &
        \multirow{2}{*}{$\phi_H$} &
        \multirow{2}{*}{$\dot{M}_{\rm acc}$ [$\dot{M}_{\rm Edd}$]} &
        \multicolumn{2}{c}{$\dot{M}_{\rm wind}$ [$\dot{M}_{\rm Edd}$]} &
        \multirow{2}{*}{$N_r \times N_{\theta} \times N_{\phi}$} &
        \multirow{2}{*}{$T_{\rm max}$ [$t_g$]} \\
        \cline{6-7}
        
        & & & & &
        $r=100\,r_g$ & $r=2000\,r_g$ & & \\
        
        \hline

        \tt{m5a0r1} & 0.0 & 5 & 40 & 21.8 & 22.2 & 61.5 & $128\times64\times32$ & 40,000 \\
        \tt{m5a0r2} & 0.0 & 5 & 46 & 45.7 & 47.2 & 135 & $128\times64\times32$ & 40,000 \\
	    \tt{m5a0r3} & 0.0 & 5 & 42 & 172 & 265 & 848 & $128\times64\times32$ & 60,000 \\
	    \tt{m5a0r4} & 0.0 & 5 & 50 & 488 & 765 & 1887 & $128\times64\times32$ & 60,000 \\
    
        \tt{m5a9r1} & 0.9 & 5 & 42 & 5.4 & 6.2 & 14.8 & $120\times72\times32$ & 30,000 \\
	    \tt{m5a9r2} & 0.9 & 5 & 40 & 16.6 & 39.2 & 74.3 & $120\times72\times32$ & 30,000 \\
	    \tt{m5a9r3} & 0.9 & 5 & 50 & 120 & 415 & 779 & $120\times72\times32$ & 30,000 \\
	    \tt{m5a9r4} & 0.9 & 5 & 52 & 414 & 1820 & 3871 & $128\times64\times32$ & 85,000 \\

	    \tt{m15a0r1} & 0.0 & 15 & 37 & 7.7 & 5.8 & 18.8 & $128\times64\times32$ & 40,000 \\
        \tt{m15a0r2} & 0.0 & 15 & 30 & 15.9 & 13.9 & 39.1 & $128\times64\times32$ & 40,000 \\
        \tt{m15a0r3} & 0.0 & 15 & 44 & 65.1 & 82.1 & 194 & $128\times64\times32$ & 40,000 \\
	    \tt{m15a0r4} & 0.0 & 15 & 44 & 144 & 193 & 439 & $128\times64\times32$ & 60,000 \\
	    \tt{m15a0r5} & 0.0 & 15 & 58 & 195 & 282 & 825 & $128\times64\times32$ & 60,000 \\
        \tt{m15a0r6} & 0.0 & 15 & 49 & 1438 & 1916 & 5468 & $120\times72\times32$ & 90,000 \\
	    \tt{m15a0r7} & 0.0 & 15 & 40 & 1592 & 2529 & 8361 & $128\times64\times32$ & 90,000 \\
        
        \tt{m15a9r1} & 0.9 & 15 & 40 & 6.8 & 7.4 & 22.0 & $128\times64\times32$ & 40,000 \\
        \tt{m15a9r2} & 0.9 & 15 & 42 & 10.2 & 19.4 & 45.4 & $128\times64\times32$ & 40,000 \\
        \tt{m15a9r3} & 0.9 & 15 & 50 & 21.4 & 64.7 & 126 & $128\times64\times32$ & 40,000 \\
        \tt{m15a9r4} & 0.9 & 15 & 55 & 37.0 & 116 & 225 & $128\times64\times32$ & 40,000 \\
	    \tt{m15a9r5} & 0.9 & 15 & 53 & 70.9 & 217 & 425 & $128\times64\times32$ & 40,000 \\
	    \tt{m15a9r6} & 0.9 & 15 & 55 & 191 & 808 & 1693 & $128\times64\times32$ & 60,000 \\
        \tt{m15a9r7} & 0.9 & 15 & 60 & 450 & 1964 & 5697 & $128\times64\times32$ & 90,000 \\
        \tt{m15a9r8} & 0.9 & 15 & 49 & 1168 & 4914 & 13025 & $128\times64\times32$ & 100,000 \\
  
        \tt{m30a0r1} & 0.0 & 30 & 34 & 14.0 & 13.6 & 39.1 & $128\times64\times32$ & 40,000 \\
        \tt{m30a0r2} & 0.0 & 30 & 44 & 158 & 175 & 392 & $120\times72\times32$ & 40,000 \\
        \tt{m30a0r3} & 0.0 & 30 & 43 & 184 & 269 & 655 & $128\times64\times32$ & 60,000 \\
	    \tt{m30a0r4} & 0.0 & 30 & 44 & 1155 & 1577 & 5238 & $128\times64\times32$ & 90,000 \\
        
	    \tt{m30a9r1} & 0.9 & 30 & 47 & 10.2 & 21.7 & 51.0 & $128\times64\times32$ & 30,000 \\
	    \tt{m30a9r2} & 0.9 & 30 & 49 & 81.0 & 256 & 529 & $128\times64\times32$ & 30,000 \\
	    \tt{m30a9r3} & 0.9 & 30 & 55 & 130 & 473 & 977 & $128\times64\times32$ & 40,000 \\
        
		\hline
        
	\end{tabular}
\end{table*}

\begin{figure}
	\includegraphics[width=\columnwidth]{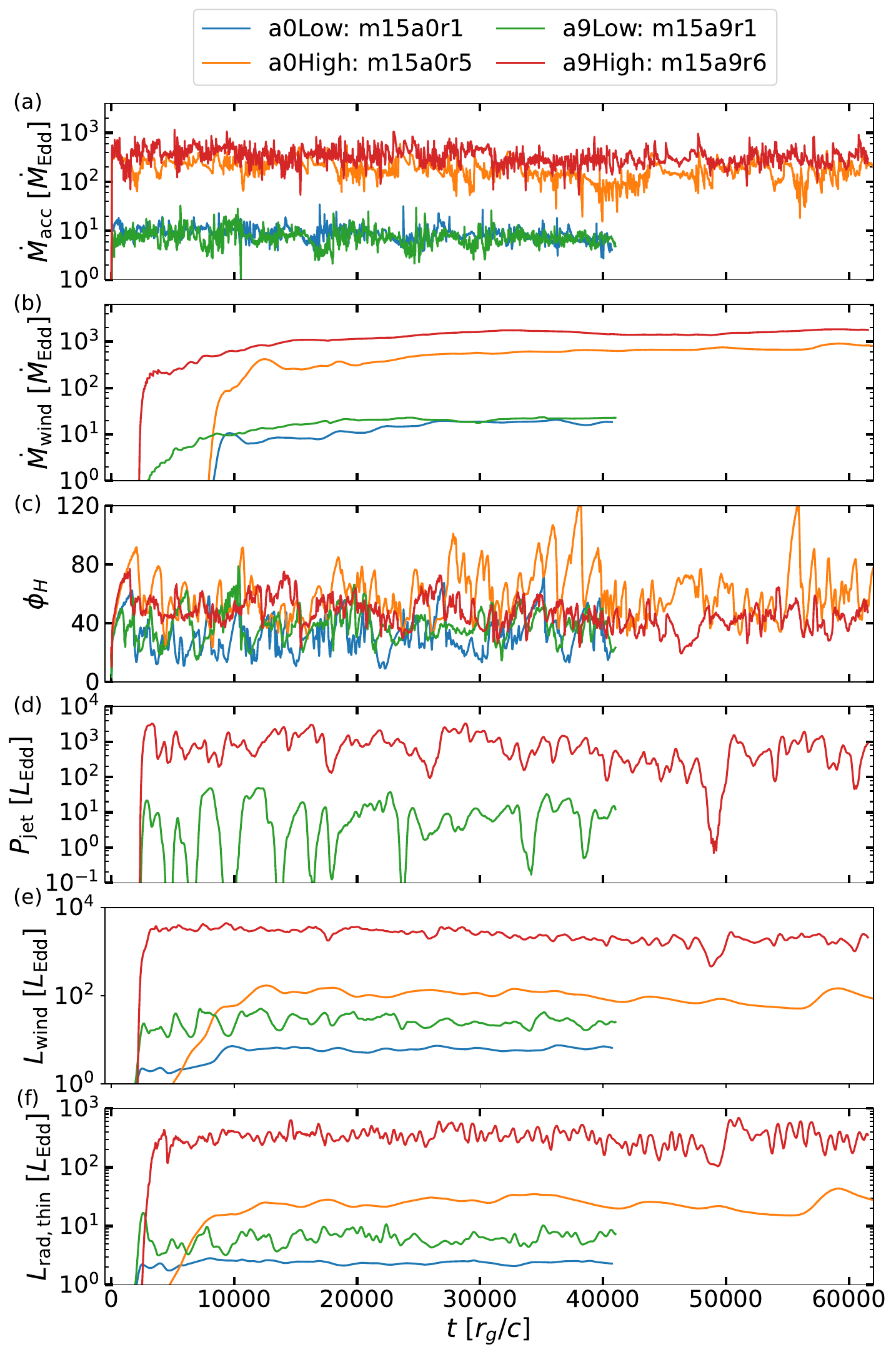}
    \vspace*{-0.2cm}
    \caption{Time evolution of accretion and outflow properties in the four fiducial models (see Table \ref{tab:selected_models}). From top to bottom panel: (a) BH mass accretion rate $\dot{M}_{\rm acc}$, (b) wind-loss rate $\dot{M}_{\rm wind}$, (c) dimensionless magnetic flux at the horizon (the MADness parameter) $\phi_H$, (d) jet total power $P_{\rm jet}$, (e) wind total power $L_{\rm wind}$, and (f) radiative luminosity $L_{\rm{}rad}$ from the outflow region outside the effective photosphere. All outflow quantities (b, d, e and f) are evaluated at $r=2000\, r_g$. For clarity, the plotted time series are sampled at intervals of $16t_g$ to reduce fast fluctuations.  }
    \label{fig:M15_time_evo}
\end{figure}

\section{Results}
\label{sec:results}

\subsection{Overview of model parameters and structure of accretion flow}
\label{sec:models_summary}

We performed 30 simulations of super-Eddington accretion flows onto stellar-mass BHs, varying three physical parameters: BH mass ($M_{\rm BH}=5$, $15$, and $30\,M_{\odot}$), BH spin ($a=0$ and $0.9$), and mass accretion rate ($\dot{M}_{\rm acc} \approx 1-2000\,\dot{M}_{\rm Edd}$). Table~\ref{tab:models} lists all models and their basic key parameters. BH masses are chosen to span the range expected for BHs in ULXs from binary population synthesis studies \citep{Wiktorowicz.2017, Misra.2023}.

\begin{table}
	\centering
        \caption{List of the four fiducial models}
        \label{tab:selected_models}
	\begin{tabular}{lcccc} 
		\hline
        \hline
		Abbreviation  & Model & $a$ & $M_{\rm BH}$ [$M_{\odot}$] & $\dot{M}_{\rm acc}$ [$\dot{M}_{\rm Edd}$]  \\ [1ex]
		\hline

		\tt{a0Low} & \tt{m15a0r1} & 0.0 & 15 & 7.7  \\
		\tt{a0High} & \tt{m15a0r5} & 0.0 & 15 & 195  \\
		\tt{a9Low} & \tt{m15a9r1} & 0.9 & 15 & 6.8  \\
		\tt{a9High} & \tt{m15a9r6} & 0.9 & 15 & 191  \\

		\hline
	\end{tabular}
\end{table}

For detailed analysis and comparison, we select four fiducial models (Table~\ref{tab:selected_models}) that span the range of spin and accretion rate: \texttt{a0Low} ($\dot{m}\sim10$, $a=0$), \texttt{a9Low} ($\dot{m}\sim10$, $a=0.9$), \texttt{a0High} ($\dot{m}\sim200$, $a=0$), and \texttt{a9High} ($\dot{m}\sim200$, $a=0.9$). These models span the parameter space and highlight the key effects of spin and accretion rate on the flow structure, outflow dynamics, and radiative output.

All simulations were evolved for at least $t = 30,000\,t_g$ (see Table~\ref{tab:models} for run times), reaching quasi-steady inflow out to $r \gtrsim 100\,r_g$ and thermally unbound outflow equilibrium out to $r \gtrsim 2000\,r_g$ in all cases. We time-average all quantities over the final $\sim10,000\,t_g$ of each simulation.

All disks have relatively high values of the MADness parameter $\phi_H$ in the quasi-steady state, with time-averaged values listed in Table~\ref{tab:models}. These accretion flows have entered the MAD state, as demonstrated by the large-amplitude variability of both $\dot{M}_{\rm acc}$ and $\phi_H$ (see Figure~\ref{fig:M15_time_evo} panels a and b), a characteristic behavior of MAD flows \citep[e.g.,][]{Tchekhovskoy.2011, McKinney.2012}. 

The accretion flow is divided into three regions for further analysis:
\begin{itemize}
    \item \textbf{Disk:} Inflowing gas with $v^r<0$;
    \item \textbf{Wind:} Outflowing gas with $v^r>0$ and magnetization $\sigma \equiv b^2/\rho < 1$\footnote{For outflows, we further distinguish between thermally bound ($-hu_t<1$) and unbound ($-hu_t>1$) outflows (dashed purple lines in Figures~\ref{fig:small-structure} and~\ref{fig:Rad-structure}). Our analysis focuses on the unbound component, as bound outflows originate from the outer disk, move very slowly ($v^r \lesssim 0.01c$), and may not reach equilibrium within our simulation time frame.}; and
    \item \textbf{Relativistic jet:} Relativistic outflowing gas ($v^r>0.6c$) with magnetically dominated energy density ($\sigma>1$), where magnetic field lines are connected to the BH.
\end{itemize}

\begin{figure}
	\includegraphics[width=\columnwidth]{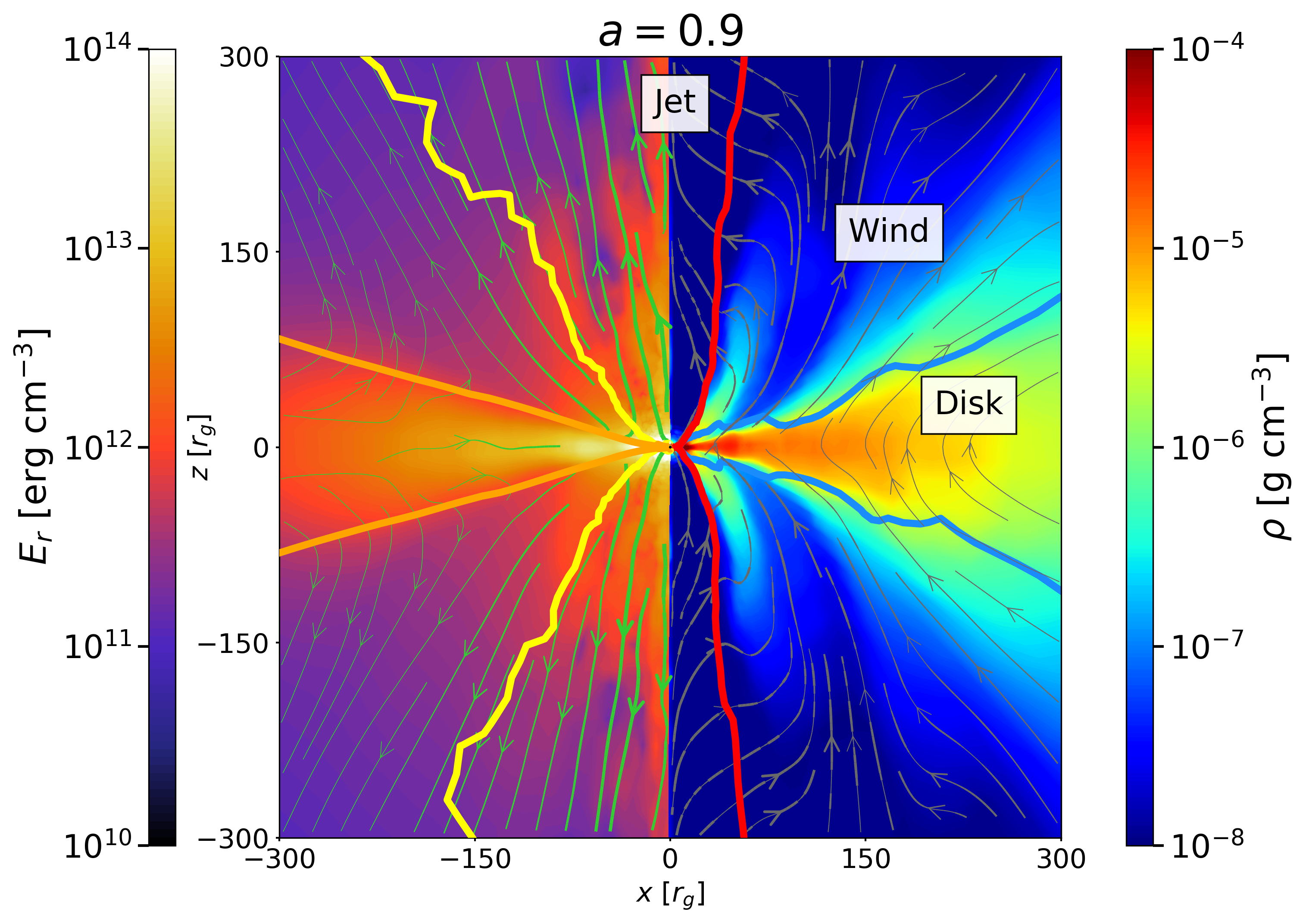}
    \includegraphics[width=\columnwidth]{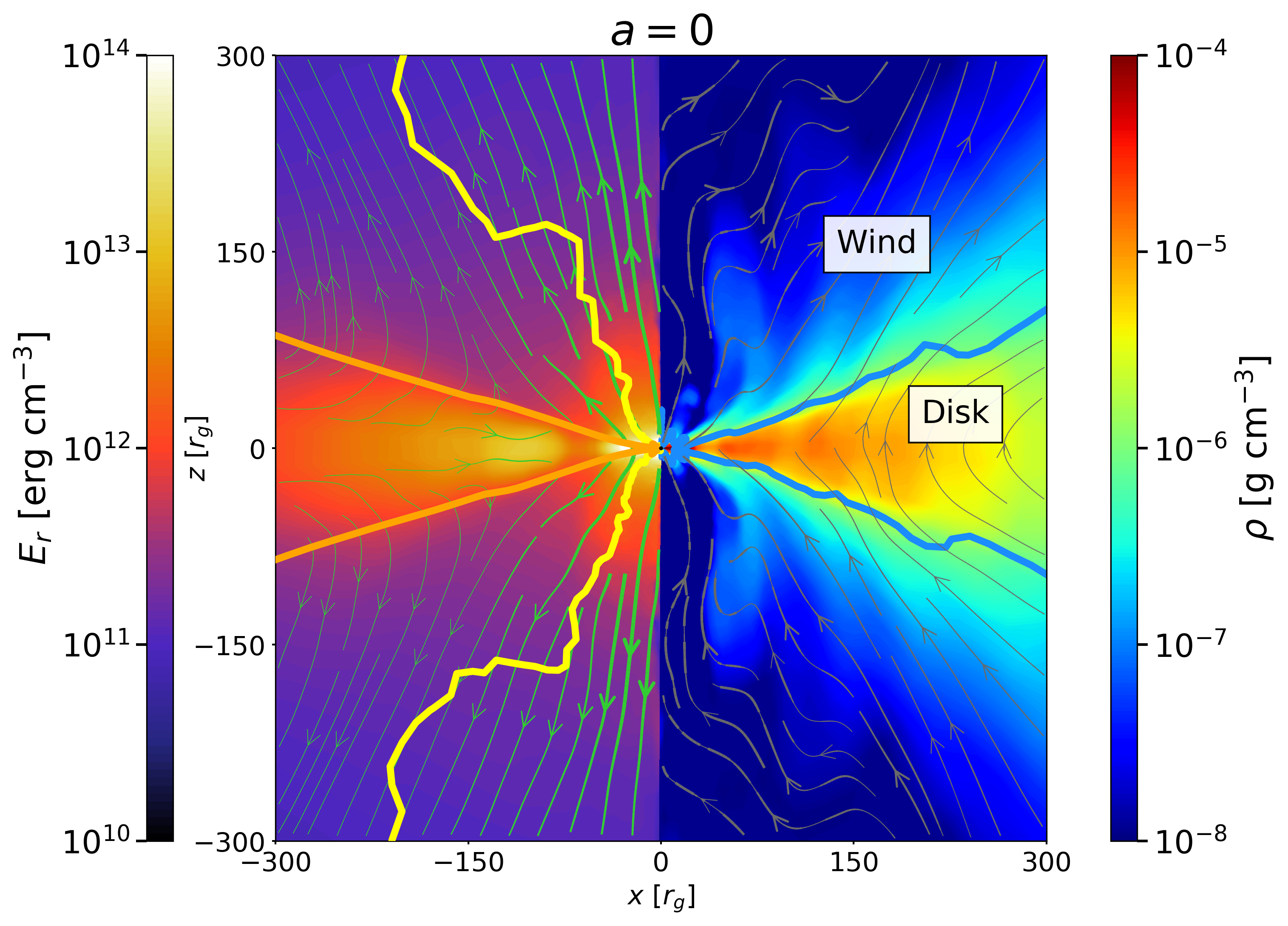}
    \vspace*{-0.2cm}
    \caption{Snapshot of a vertical slice through the inner region of model \texttt{m15a9r1} (top) and model \texttt{m15a0r1} (bottom) at $t=40000\, t_g$. The left half shows the radiation energy density $E_r$ (units of erg cm$^{-3}$) and the right half shows the gas density $\rho$ (units of g cm$^{-3}$). The blue contours mark the inflow-outflow boundary ($v^r=0$), and the red contours mark the jet-wind boundary ($\sigma=1$). The yellow and orange contours indicate the electron-scattering photosphere ($\tau_{\rm es}=1$) and the effective photosphere ($\tau_{\rm eff}=1$), respectively.
    Grey lines show magnetic field lines, and green lines show the radiation flux; line thickness indicates the field strength for magnetic field lines or the flux magnitude for radiation streamlines. }
     
    \label{fig:snapshot}
\end{figure}

Figure~\ref{fig:snapshot} shows vertical slices through the inner regions of models \texttt{m15a9r1} (top) and \texttt{m15a0r1} (bottom) at $t=40000\,t_g$. 
The right panels show the gas rest-mass density $\rho$, with the blue contour marking the inflow-outflow boundary ($v^r=0$) and the red contour marking the jet-wind boundary ($\sigma=1$). The left panels show the radiation energy density in the radiation rest frame $E_r$, with the yellow and orange contours indicating the electron-scattering photosphere ($\tau_{\rm es}=1$) and the effective photosphere ($\tau_{\rm eff}=1$), respectively.

Strong magnetic field lines (grey lines in the right panels of Figure~\ref{fig:snapshot}) thread the inner disk, wind, and, if BH spins fast, jet. The radiation flux streamlines (green lines in the left panels) are consistent with most of the escaping radiation originating in the accretion flow. The radiation broadly follows the disk inflow before turning outward and propagating approximately radially with the wind. This behavior occurs because the inner disk and wind are optically thick, so that radiation is trapped and advected with the gas.

It is worth noting that under our classification, the wind also includes the jet sheath, which is the relatively dense, moderately relativistic layer surrounding the magnetically dominated jet core. This matter is supplied primarily by the disk wind and subsequently accelerated through its interaction with the inner relativistic jet. Because our distinction between the wind and jet is based on the magnetization threshold $\sigma=1$, the transition between these components is continuous rather than a sharp physical boundary. A more detailed analysis of the accretion flow structure and its dependence on key physical parameters is presented in Section~\ref{sec:structure}.

\subsection{Energy carried by wind and jet}
\label{sec:luminosity}

We focus on the energy carried by the wind and jet, through which energy can escape from the accretion flow and provide observational signatures. In particular, the radiative, kinetic, and electromagnetic luminosities quantify the main channels through which the flow transports energy outward.

The wind and jet carry different mixtures of these energy components at different radii. 
We find that the thermally unbound wind, at large radii, carries most of its energy as kinetic power, although the radiative and EM components can also contribute significantly. By contrast, the relativistic jet, when present, remains magnetically dominated and carries most of its power in EM form. We therefore analyze the radiative, kinetic, and EM powers of the wind separately in Section~\ref{sec:wind_energy}, while characterizing the jet by its total power (primarily electromagnetic) in Section~\ref{sec:jet_energy}. We also discuss the conversion between different energy components in the outflow at different radii in Section~\ref{sec:energy_flux}. 

Numerical values of the wind luminosities and efficiencies for all models, together with the jet properties of the $a=0.9$ models, are listed in Appendix~\ref{ap:table}.

\begin{figure*}
    \centering
    \includegraphics[width=0.9\textwidth]{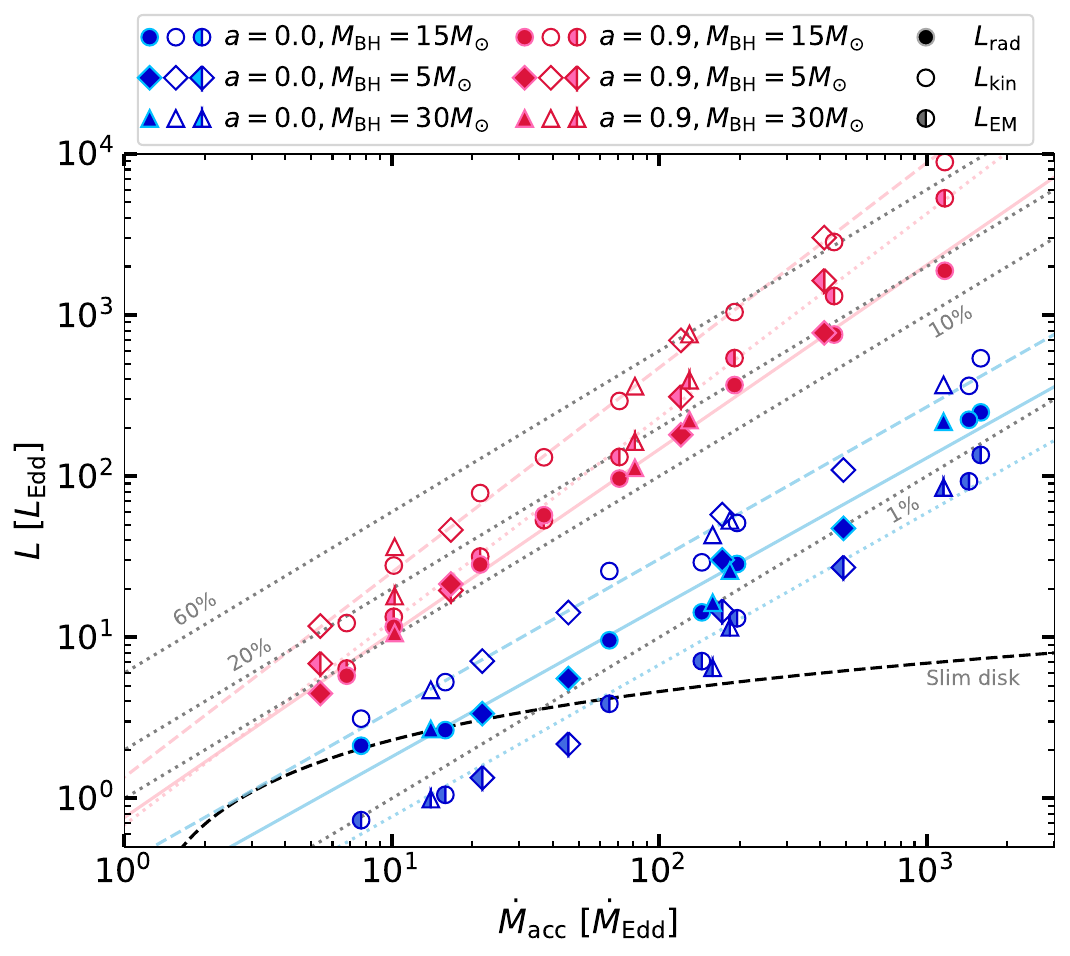}
    \vspace*{-0.22cm}
    \caption{Radiative luminosity $L_{\rm rad}$ (fully filled symbols), kinetic luminosity $L_{\rm kin}$ (open symbols) and EM luminosity $L_{\rm EM}$ (half-filled symbols) of the thermally unbound wind. All quantities are measured at $r = 2000\,r_g$ and expressed in units of Eddington luminosity $L_{\rm Edd}$. For $L_{\rm rad}$, we include only the wind outside the effective photosphere ($\tau_{\rm eff}<1$). Blue and red symbols denote simulations with BH spin $a = 0$ and $a = 0.9$, respectively. Diamonds, circles, and triangles correspond to BH masses of $5$, $15$, and $30\,M_\odot$. 
    The colored solid lines show fits to $L_{\rm rad}$, dashed lines show fits to $L_{\rm kin}$, dotted lines show fits to $L_{\rm EM}$. Grey dotted lines indicate constant energy efficiencies of $1\%$, $10\%$, $20\%$, and $60\%$, while the black dashed curve shows the slim-disk luminosity as a function of $\dot{M}_{\rm acc}$.}

    \label{fig:luminosity}
\end{figure*}

\subsubsection{Energy output from winds}
\label{sec:wind_energy}

We first examine the energetics of the thermally unbound wind. 
Figure~\ref{fig:luminosity} shows the time-averaged kinetic luminosity $L_{\rm kin}$, radiative luminosity $L_{\rm rad}$,  and EM luminosity $L_{\rm EM}$ as functions of $\dot{M}_{\rm acc}$ for all models. All quantities are measured at $r=2000\,r_g$, with the jet region excluded. Each model produces three data points: full-filled symbols denote $L_{\rm rad}$, open symbols denote $L_{\rm kin}$, and half-filled symbols denote $L_{\rm EM}$.
For $L_{\rm rad}$, we include only the thermally unbound wind region outside the effective photosphere ($\tau_{\rm eff}<1$, see Equation~\ref{eq:thin_rad_L}).
For comparison, the grey dotted lines indicate constant energy efficiencies, while the black dashed line shows the slim disk prediction for radiative luminosity $L_{\rm rad}/L_{\rm Edd} \simeq 1 + \ln \dot{m}$ \citep{Abramowicz.1988}.

Several clear trends emerge from the figure:

\begin{itemize}
    \item \textbf{The wind luminosities show no sign of Eddington regulation.} Instead of increasing logarithmically with $\dot{M}_{\rm acc}$, the radiative, kinetic, EM, and total luminosities of the wind follow approximate power-law scalings. For the $a=0.9$ models, the luminosities grow faster than linearly with accretion rate, indicating that energy output becomes even more efficient at higher Eddington ratios.
    
    \item \textbf{In all models, kinetic energy dominates over both radiative and EM energy in the wind at large radii.} We find $L_{\rm kin} > L_{\rm rad}$ and $L_{\rm kin} > L_{\rm EM}$ across the full parameter range, with $L_{\rm kin}$ often exceeding the sum of the other two components, indicating that the wind carries most of its energy as bulk kinetic power at $r=2000 r_g$. The conversion between kinetic, EM, and radiative energy in the outflow as it propagates outward is further examined in Section~\ref{sec:energy_flux}.
    
    \item \textbf{BH spin substantially boosts the wind energy output.} At fixed $\dot{M}_{\rm acc}$, the $a=0.9$ models (red symbols) produce larger $L_{\rm kin}$, $L_{\rm rad}$,  and $L_{\rm EM}$ than the $a=0$ models (blue symbols). The kinetic, radiative, and EM efficiencies remain below a few percent for $a=0$, but reach tens of percent for $a=0.9$. The EM component is negligible in the $a=0$ models, but becomes larger than the radiative component in the $a=0.9$ models.

    \item \textbf{BH mass has no systematic effect over the stellar-mass range studied here.} Models with $M_{\rm BH}=5$, $15$ and $30\,M_\odot$ (diamonds, circles, and triangles, respectively) follow the same luminosity trends.

    \item \textbf{Context among previous simulations:} Figure~\ref{fig:literature} places the magnitude of published efficiencies in context. In particular, high-BH-spin MAD models in the literature tend to reach radiative efficiencies that are large, compared to non-MAD models. Our simulation results are consistent with previous MAD simulations and show higher radiative and kinetic efficiencies than most previous non-MAD simulations for both BH spins. This demonstrates that MAD disks are more efficient at extracting energy. However, we caution that direct comparisons across simulations should be made carefully, as our setup differs in various aspects, including initial conditions, physical prescriptions, and the radius at which we evaluate wind properties, which is significantly larger than in most prior studies. We further discuss these comparisons in Section~\ref{sec:literature}.

\end{itemize}

Although part of the thermally unbound wind at high inclinations and large radii can be contaminated by the outflow originating from the outer non-equilibrium region, most of the wind energy flux, including the radiative, kinetic, and EM components, is located at low inclinations, where they are connected to the quasi-steady inner disk. It allows us robust measurements of wind fluxes far from the BH.

We fit the wind radiative, kinetic and EM luminosities as power-law functions of the BH accretion rate, valid over $1 < \dot{m} < 2000$. 
For $a=0$, we obtain 
\begin{align}
    \label{eq:luminosity_a0_kin}
    L_{\rm kin}/L_{\rm Edd} &= 0.39\,\dot{m}^{0.95},\\
    \label{eq:luminosity_a0_rad}
    L_{\rm rad}/L_{\rm Edd} &= 0.20\,\dot{m}^{0.94},\\
    \label{eq:luminosity_a0_em}
    L_{\rm EM}/L_{\rm Edd} &= 0.07\,\dot{m}^{1.00}.
\end{align}
For $a = 0.9$, we obtain
\begin{align}
    \label{eq:luminosity_a9_kin}
    L_{\rm kin}/L_{\rm Edd} &= 1.51\,\dot{m}^{1.25}, \\
    \label{eq:luminosity_a9_rad}
    L_{\rm rad}/L_{\rm Edd} &= 0.79\,\dot{m}^{1.13}, \\
    \label{eq:luminosity_a9_em}
    L_{\rm EM}/L_{\rm Edd} &= 0.69\,\dot{m}^{1.26}.
\end{align}
These fits show that the $a=0$ models have nearly constant wind efficiencies, while the $a=0.9$ models have the super-linear exponents, pointing to a spin-dependent mechanism that becomes more effective at higher Eddington ratios. 
In the following sections, we show that these scalings arise from the combined effects of wind mass loading, wind velocity, magnetic energy extraction, and photon escape through the polar funnel, which jointly govern how much mass and energy are carried away from the system by the wind.

\subsubsection{Energy output from relativistic jets}
\label{sec:jet_energy}

\begin{figure}
	\centering
    \includegraphics[width=\columnwidth]{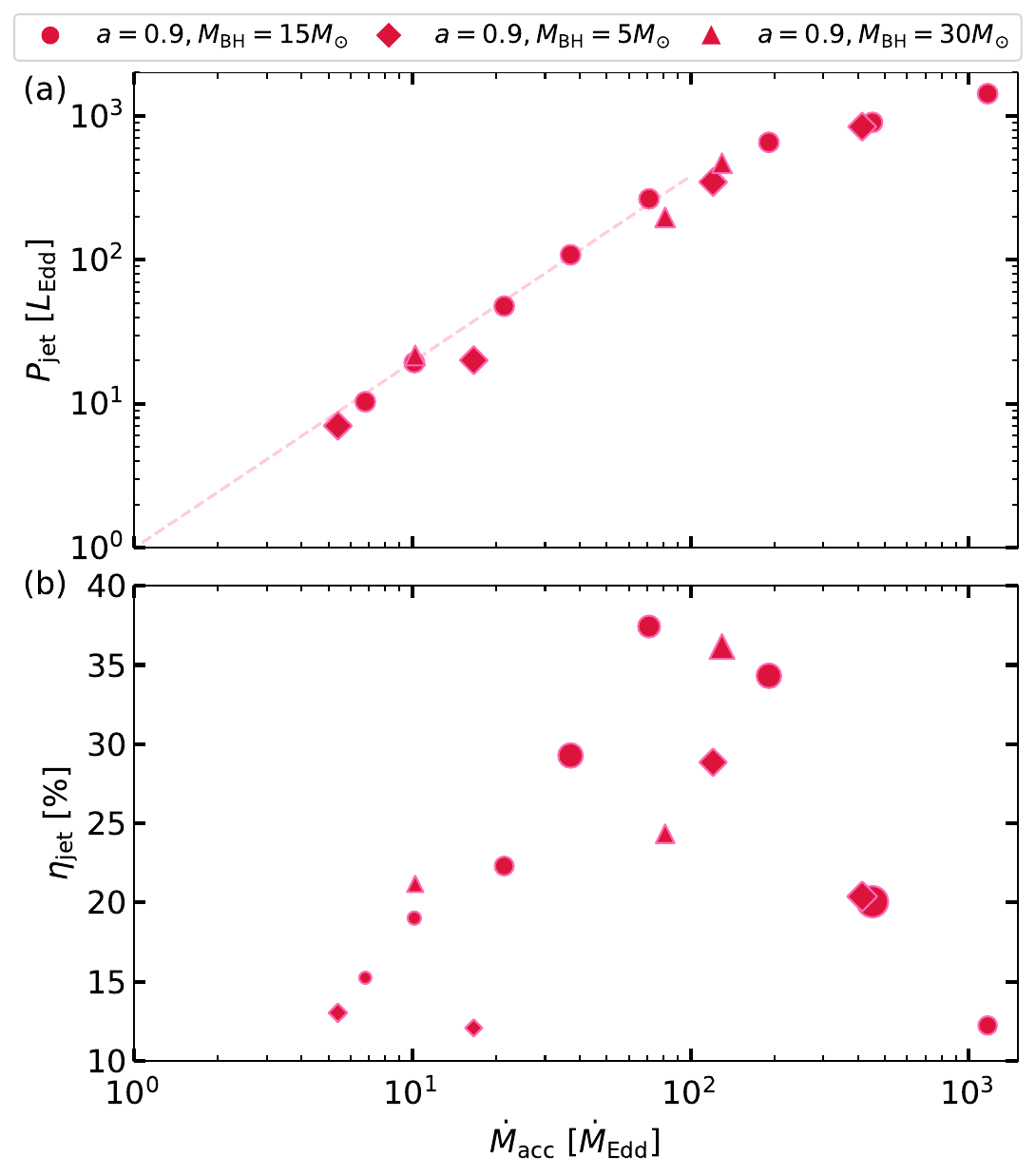}
    \vspace*{-0.2cm}
    \caption{ Jet power and jet efficiency for the $a=0.9$ models. Panel (a) shows the time-averaged total jet power $P_{\rm jet}$, measured at $r=2,000 r_g$ and expressed in units of Eddington luminosity $L_{\rm Edd}$, as a function of $\dot{M}_{\rm acc}$. The dashed line shows the best power-law fit to $P_{\rm jet}$ for $1<\dot{m}<100$. Panel (b) shows the corresponding jet efficiency $\eta_{\rm jet}$. The marker size scales with the dimensionless magnetic flux through the horizon $\phi_H$ relative to its maximum among the $a=0.9$ models. $P_{\rm jet}$ increases with $\dot{M}_{\rm acc}$ below $\dot{m} \sim 100$, but saturates at higher $\dot{m}$. $\eta_{\rm jet}$ correspondingly turns over and declines in the hyper-Eddington regime.}
    \label{fig:jet_power}
\end{figure}

\label{sec:MAD_magnetization}
\begin{figure}
    \centering
	\includegraphics[width=0.96\columnwidth]{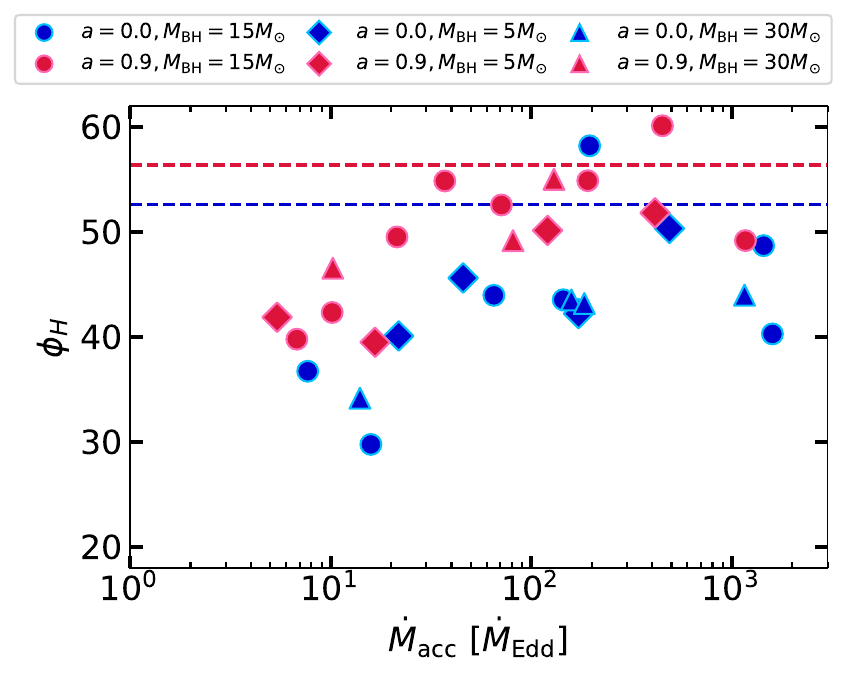}
    \vspace*{-0.2cm}
    \caption{Dimensionless magnetic flux threading the BH horizon $\phi_H$, as a function of mass accretion rate $\dot{M}_{\rm acc}$ for all simulations. Blue and red dashed lines indicate the saturated values found in nonradiative MAD simulations for $a=0$ and $a=0.9$, respectively. The value of $\phi_H$ increases with $\dot{m}$ over $1 \lesssim \dot{m} \lesssim 100$, reaches a maximum around $\dot{m}\sim100$, and declines at the highest accretion rates.}
    \label{fig:mad_parameter}
\end{figure}

When the BH spin is high and sufficient magnetic flux is accumulated near the horizon, the Blandford-Znajek mechanism \citep{Blandford.1977} launches a powerful relativistic jet. This jet provides an additional energy output beyond the wind. It extracts energy from the spinning BH, accelerating and transporting gas to large scales while carrying substantial EM power. Moreover, the jet evacuates the polar region, carving out an optically thin funnel that allows photons to escape freely and boosts the radiative output.

Figure~\ref{fig:jet_power} panel (a) shows the time-averaged total jet power $P_{\rm jet}$, measured at $r=2000\,r_g$, as a function of $\dot{M}_{\rm acc}$ for the $a=0.9$ models. While $P_{\rm jet}$ includes all energy components, it is dominated by EM energy, as expected for a magnetically dominated jet.

The jet power increases approximately as a power law for $1<\dot{m}\lesssim100$, with only a weak dependence on $M_{\rm BH}$ over $5-30\,M_\odot$. We obtain the best-fit relation: 
\begin{equation}
    P_{\rm jet}/L_{\rm Edd} = 0.86\,\dot{m}^{1.29}  \qquad (1 < \dot{m} \lesssim 100).
\end{equation}
At $\dot{m}\gg100$, the relation flattens and becomes sub-linear.

The flattening of $P_{\rm jet}$ at very high $\dot{m}$ is connected to the magnetic flux threading the BH horizon. The dimensionless MADness parameter $\phi_H$ (Equation~\ref{eq:mad}) quantifies the magnetization of the inner accretion flow. Figure~\ref{fig:mad_parameter} shows $\phi_H$ as a function of $\dot{M}_{\rm acc}$ for all simulations. For $1<\dot{m}\lesssim100$, $\phi_H$ increases with $\dot{m}$. A larger inflow rate raises the disk scale height. It allows the innermost disk to confine stronger magnetic flux. This trend is qualitatively consistent with the recent GRRMHD simulations of \citet{Ricarte.2023}.

The $\phi_H$ reaches its maximum around $\dot{m}\sim100$ and declines at much higher accretion rates. Around this transition, $\phi_H$ approaches the saturation levels measured in non-radiative MAD simulations \citep{Narayan.2022}. 
At higher $\dot{m}$, the absolute magnetic flux threading the horizon, $\Phi_H$, grows more slowly than $\sqrt{\dot{M}_{\rm acc}}$, causing the dimensionless flux $\phi_H$ to decline. The physical origin of this flux saturation likely relates to the ability of the inner disk to confine vertical magnetic flux at very high accretion rates, which warrants further investigation.

Figure~\ref{fig:jet_power} panel (b) further shows the jet total energy efficiency $\eta_{\rm jet}$ (Equation \ref{eq:eta_jet}) as a function of $\dot{M}_{\rm acc}$, with the marker size indicating the value of $\phi_H$ relative to its maximum among the $a=0.9$ models. $\eta_{\rm jet}$ initially increases with $\dot{m}$ while $\phi_H$ continues to grow. Once $\phi_H$ reaches its maximum and begins to decline, the jet power grows more slowly than the accretion-energy supply, causing $\eta_{\rm jet}$ to turn over and decrease at the highest accretion rates.

\begin{figure*}
	\centering
    \vspace*{-0.1cm}
    \includegraphics[width=\textwidth]{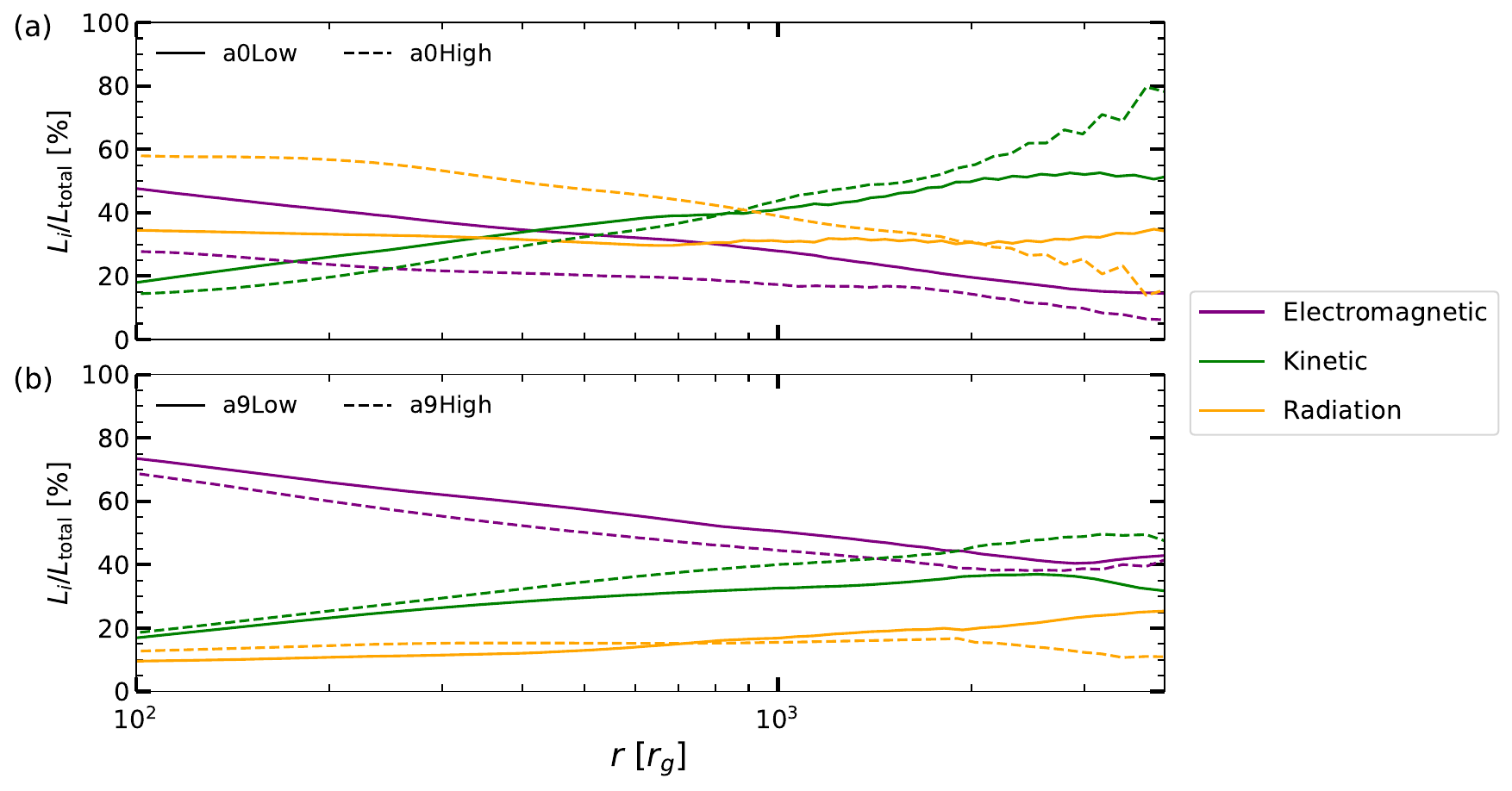}
    \vspace*{-0.3cm}
    \caption{ Radial redistribution of the energy fluxes carried by the thermally unbound outflow (wind + jet) in the four fiducial models. Purple, green, and orange curves denote the fractional contribution of electromagnetic power $L_{\rm EM}$, kinetic luminosity $L_{\rm kin}$, and bolometric radiative luminosity $L_{\rm rad}$ to the total power $L_{\rm total}$, respectively, with the three components summing to unity at each radius. Panel (a) shows the $a = 0$ models (\texttt{a0Low} as solid curves, \texttt{a0High} as dashed curves), while panel (b) shows the $a = 0.9$ models (\texttt{a9Low} as solid curves, \texttt{a9High} as dashed curves). The declining EM fraction and rising kinetic fraction show that electromagnetic energy is progressively converted into kinetic energy as the outflow expands. }
    \label{fig:wind_power_components}
\end{figure*}

\subsubsection{Energy conversion in the outflow}
\label{sec:energy_flux}

As the outflow (wind and jet) propagates outward, a significant fraction of its energy is converted from one form to another, particularly inside the photosphere.

Figure~\ref{fig:wind_power_components} shows the radial profiles of the fractional contributions of kinetic (green),  radiative (orange), and EM (purple) power to the total outflow power $L_{\rm total}$ for the four fiducial models (panel a: $a=0$; panel b: $a=0.9$). In all cases, EM and radiative energy fractions decline with radius, while the kinetic fraction increases, indicating a progressive conversion of radiation and magnetic field energy into wind kinetic energy. At fixed spin, the relative energy partition is largely independent of the accretion rate.

In the $a = 0$ models, no relativistic jet is present, and the EM energy carried by the outflow is small. Radiation carries the largest fraction of the total power at small radii. As the dense wind expands, trapped radiation performs work on the gas, converting radiative energy into bulk kinetic energy. This conversion is more efficient at higher accretion rates (dashed lines), where the denser wind remains optically thick to larger radii.

In the $a = 0.9$ models, the relativistic jet carries a large electromagnetic flux, particularly at small radii ($r \lesssim 500\,r_g$), where EM power dominates the total energy budget. As the jet and magnetized wind propagate outward, the EM energy fraction declines and the kinetic energy fraction rises, indicating that Poynting flux is dissipated and transferred to accelerating the gas. 
By $r \sim 2000\,r_g$, the radius at which we evaluate the wind energetics in Figure~\ref{fig:luminosity}, kinetic energy becomes the dominant component.

Our results are qualitatively consistent with \citet{Yang.2023}, who studied a super-Eddington MAD simulation \citep{Thomsen.2022} using a virtual-particle approach \citep{Yuan.2015} and found that Poynting flux is progressively converted into kinetic energy as the outflow expands.

These energy-conversion processes explain why kinetic power dominates the energy output in all our models, while many previous super-Eddington simulations reported that radiation energy dominates outflows. In many of those studies, the accretion flows remained in the SANE state, which generally produced less energetic winds. More importantly, differences in measurement radius and photosphere treatment also contribute to the discrepancy. At smaller radii, a larger fraction of the energy remains in radiative or electromagnetic form, whereas our extended domain captures its subsequent conversion into kinetic energy. This highlights the importance of using a sufficiently extended simulation domain and evolving the flow until outflow equilibrium is reached over a large region, so that energy outputs can be evaluated outside the photosphere, where the outflow has completed most of its energy conversion and the true asymptotic energy budget can be reliably measured.

\begin{figure*}
    \centering
    \includegraphics[width=0.9\textwidth]{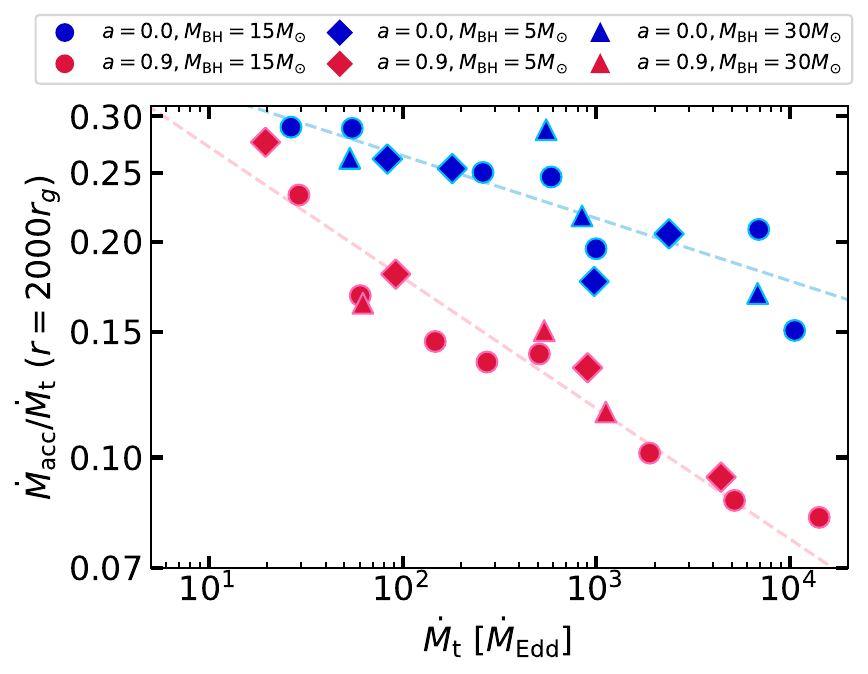}
    \vspace*{-0.2cm}
    \caption{Ratio of BH accretion rate to total mass-supply rate, $\dot{M}_{\rm acc}/\dot{M}_{\rm t}$, as a function of $\dot{M}_{\rm t}$. The total mass-supply is defined as $\dot{M}_{\rm t}=\dot{M}_{\rm acc}+\dot{M}_{\rm wind}$, where the wind rate is measured for thermally unbound wind at $r = 2000 r_g$. Different symbols represent BH masses ($5$, $15$, and $30 M_\odot$); colours distinguish BH spins ($a = 0$ in blue, $a = 0.9$ in red). Colored dashed lines indicate best-fit power-law relations. The ratio decreases with increasing mass supply rate and is systematically lower for $a = 0.9$ than for $a = 0$.  }
    \label{fig:accretion_ratio}
\end{figure*}

\subsection{Accretion versus ejection ratio}
\label{sec:accretion_ratio}

The fraction of gas that reaches the BH, rather than being ejected in winds, is a key quantity for understanding accretion feedback, the mass budget of super-Eddington systems, and the connection to binary evolution and SMBH growth. We characterize this by the mass accretion ratio $\dot{M}_{\rm acc}/\dot{M}_{t}$, where the total mass-supply rate $\dot{M}_{t} \equiv \dot{M}_{\rm acc} + \dot{M}_{\rm wind}$ is the sum of the time-averaged BH accretion rate and wind loss rate.

Figure~\ref{fig:accretion_ratio} shows the accretion ratio $\dot{M}_{\rm acc}/\dot{M}_{t}$ as a function of $\dot{M}_{t}$ for all simulations where $\dot{M}_{\rm wind}$ is measured for thermally unbound winds measured at $r=2000\,r_g$, consistently with previous energy flux analysis. Several features stand out. First, the BH accretes only around $7\%$ to $30\%$ the supplied gas from large radii; the remainder is ejected in winds. Second, for a given $\dot{M}_{t}$, the accretion ratio is lower for $a=0.9$ than for $a=0$, indicating that high BH spin enhances mass loss. This is consistent with our earlier finding that high spin produces stronger kinetic-energy feedback (Section~\ref{sec:luminosity}), thereby driving more gas away from the system. Third, the ratio decreases with increasing $\dot{M}_{t}$ as a power law, with no obvious dependence on BH mass ($5-30\,M_{\odot}$). This is likewise consistent with our finding that kinetic power increases with $\dot{M}_{\rm acc}$ (Section~\ref{sec:luminosity}): higher mass supply rates drive stronger winds, which carry away a larger fraction of the gas.

We fit the following power-law relations for $10 \lesssim \dot{M}_{t}/\dot{M}_{\rm Edd} \lesssim 10^4$:
\begin{align}
    \dot{M}_{\rm acc}/\dot{M}_{t} \ (r=2000\, r_g) &= 
    \begin{cases}
        0.40\,\bigl(\dot{M}_{t}/\dot{M}_{\rm Edd}\bigr)^{-0.09}, & a = 0, \\[0.2cm]
        0.41\,\bigl(\dot{M}_{t}/\dot{M}_{\rm Edd}\bigr)^{-0.18}, & a = 0.9.
    \end{cases}
    \label{eq:mass_ratio_fits}
\end{align}
The steeper decline for $a=0.9$ reflects the greater wind-launching efficiency in rapidly spinning systems, which becomes more prominent at higher Eddington ratios. Over the range $\dot{M}_{t}/\dot{M}_{\rm Edd}=10$ to $10^4$, the accretion ratio decreases from $\sim30\%$ to $\sim15\%$ for $a=0$, and from $\sim 30\%$ to $\sim7\%$ for $a=0.9$.

\begin{figure*}
    \centering
    \includegraphics[width=0.9\textwidth]{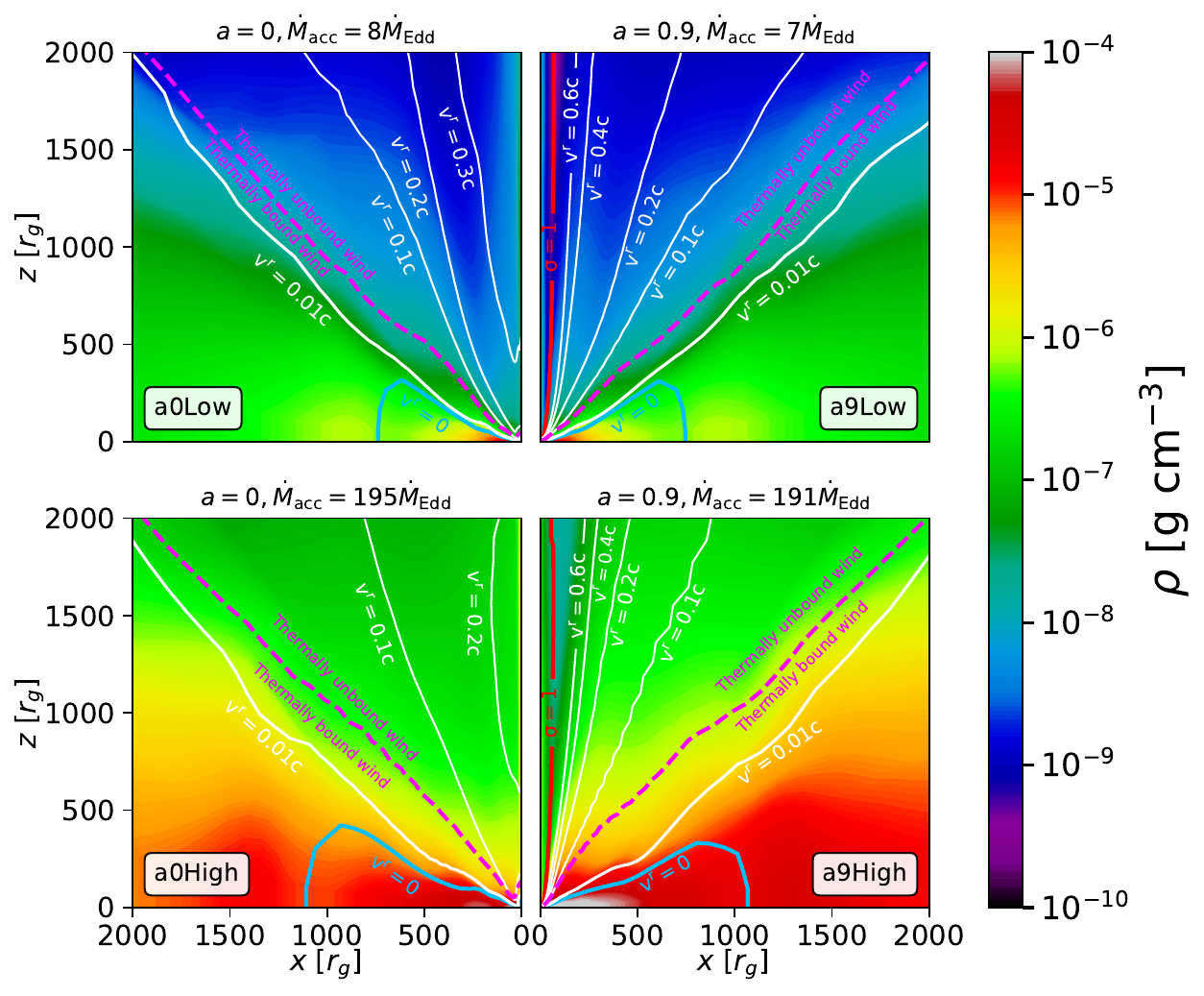}
    \vspace*{-0.2cm}
    \caption{Time- and azimuthally-averaged 2D profiles of rest-mass density $\rho$ for the four fiducial models, shown within $r = 600 r_g$. Blue contours mark the inflow-outflow boundary ($v_r = 0$), red contours denote the jet boundary ($\sigma = 1$), pink dashed lines show the thermally bound-unbound wind boundary ($-hu_t = 1$), and white lines represent contours of constant radial velocity $v_r$. Spinning BH models (\texttt{a9Low} and \texttt{a9High}, right panels) exhibit relativistic jets, while non-spinning models (\texttt{a0Low} and \texttt{a0High}, left panels) show slower polar outflows.
    }

    \label{fig:small-structure}
\end{figure*}

An important limitation of our wind mass-loss estimates is that the disk has reached inflow equilibrium only within $r \lesssim 100 r_g$, while the outer disk is still evolving from the initial conditions. Consequently, turbulence and wind production at larger radii has not be fully developed within the relatively short simulation time. It is therefore difficult to know whether the wind mass-loss rate measured at $r=2000r_g$ may therefore overestimate or underestimate the total mass loss expected from a fully developed accretion flow. For comparison, we also evaluate the conservative lower limit of the total outward mass flux at $r=100r_g$, within the inflow-equilibrium region, in Appendix~\ref{ap:launching}. Although this measurement includes all outward-moving gas rather than only the thermally unbound component, it has qualitatively similar trends with BH spin and mass supply rate, with accretion ratios approximately twice those inferred at $r=2000r_g$.

\subsection{Dependence of the accretion flow structure and dynamics on BH parameters}
\label{sec:structure}

We now examine how BH spin and accretion rate shape the structure and dynamics of the accretion flow, following our analysis of the global outflow energetics.

Figure~\ref{fig:small-structure} shows the time- and $\phi$-averaged vertical slices of the gas density and velocity structure within $r=600\,r_g$ for each of the four fiducial models (see Table~\ref{tab:selected_models}). In each panel, the accretion flow is divided into three components: the disk, the wind, and, when present, the relativistic jet.
The boundaries between these regions are defined as in Figure~\ref{fig:snapshot}: red contours mark the jet-wind boundary ($\sigma=1$), blue contours mark the inflow-outflow boundary ($v^r=0$), and white lines show constant $v^r$.

The overall structure is consistent with previous numerical studies of super-Eddington disks \citep[e.g.,][]{Jiang.2014, Sadowski.2016, Dai18, Yoshioka.2022}. A geometrically thick disk forms along the equatorial plane due to large radiation and magnetic pressure. The high-pressure disk launches a wide-angle, anisotropic wind with lower-density, high-velocity ($v^r>0.1c$) ultrafast outflows (UFOs) at low inclinations and denser, slower winds at high inclinations. The mean flow geometry associated with wind launching from the disk surface are examined further in Appendix \ref{ap:launching}. These broad features appear in all four models, while BH spin and accretion rate mainly modify the polar funnel, outflow velocity, and gas density.

\subsubsection{Effect of the BH spin}
Figure~\ref{fig:small-structure} shows that the flow structure at low inclinations is highly sensitive to BH spin. In the $a=0.9$ models, a relativistic jet forms via the Blandford-Znajek mechanism \citep{Blandford.1977}. Even outside the magnetically dominated jet region, the low-inclination wind reaches substantially higher velocities than in the $a=0$ models: $v^r \sim0.4-0.5c$ for $a=0.9$, compared with $\lesssim0.3c$ for $a=0$. 

When $a=0$, after the disk wind is launched and accelerated through Blandford-Payne mechanism \citep{Blandford.1982}, part of the wind is gradually directed toward the pole and becomes collimated at large radii. As the magnetized wind rotates, the poloidal magnetic field is wound into a strong, large-scale toroidal component, $B_{\phi}$. The associated magnetic tension acts inward on the gas enclosed by the toroidal field loops and collimates the flow toward the rotational axis.

\begin{figure}
    \includegraphics[width=\columnwidth]{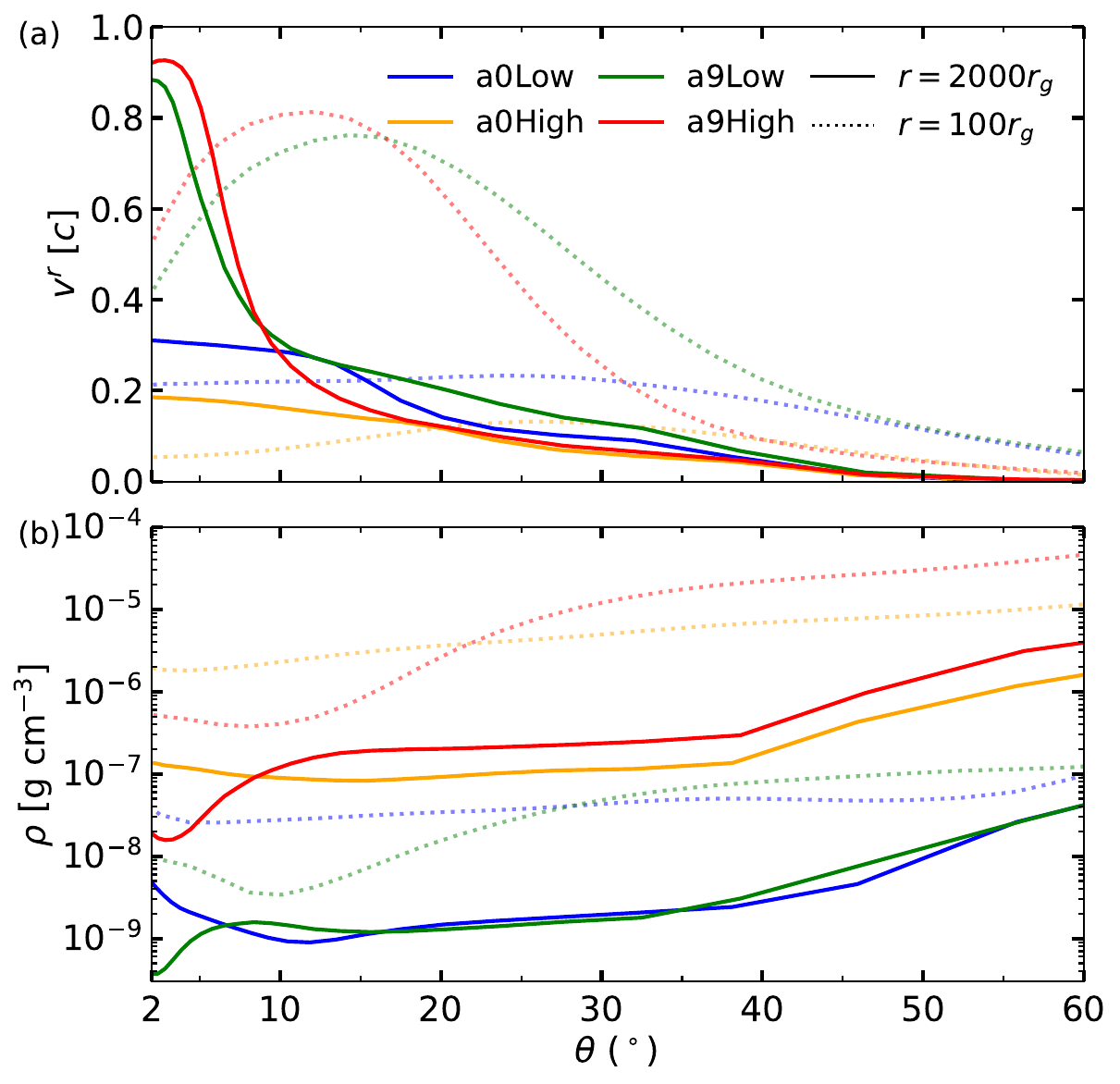}
    \vspace*{-0.4cm}
    \caption{Time- and $\phi$-averaged polar profiles of the gas on a sphere of radius $r = 2000r_g$ (solid lines) and $100r_g$ (dotted lines) for the four fiducial models: 
    (a) projected radial velocity $v^r$ of the gas in units of $c$, (b) gas density $\rho$ in units of g cm$^{-3}$.  }
    \label{fig:angular}
\end{figure}

Figure~\ref{fig:angular} further shows the corresponding angular profiles of the wind radial velocity (panel a) and gas density $\rho$ (panel b) at $r=2000\,r_g$ (solid lines) and $100\,r_g$ (dotted lines). At both radii, the wind velocity remains much higher at low inclinations for $a=0.9$ than for $a=0$, reflecting the influence of the relativistic jet and the surrounding magnetized wind on the polar flow structure. In the $a=0.9$ models, the gas density drops near the polar axis as the jet helps evacuate gas. 

The spin dependence becomes confined to much smaller polar angles as the outflow propagates outward. 
At $r=100\, r_g$, differences between the two spin models are apparent out to $\theta\gtrsim 30-40^\circ$, whereas by $r=2000\, r_g$, they are largely confined to $\theta\gtrsim 10^\circ$. This reflects the strong collimation of the relativistic jet and the surrounding fast polar outflow. At larger inclinations, both spin models show similarly slow and dense winds.

This spin-dependent flow structure directly explains our earlier findings (see Sections~\ref{sec:luminosity} and~\ref{sec:accretion_ratio}) that high-spin models produce significantly higher kinetic luminosities and more efficient mass ejection in the wind: the polar wind is much faster in the $a=0.9$ models.  

In the $a=0.9$ models, relativistic jets are produced, which help evacuate the polar funnel. However, despite these jet-mediated effects, the wind remains the dominant carrier of mass and kinetic energy in the outflow. Even in the $a=0.9$ models, the wind carries most of the ejected gas and a substantial fraction of the total kinetic power.  The jet's primary contribution to the kinetic power is indirect; it supplies electromagnetic energy that can be transferred to the surrounding wind, helping to accelerate it and boosting its kinetic power. This additional energy input, combined with the higher gas densities and faster velocities in the polar wind region, explains why high-spin models produce significantly higher kinetic luminosities and more efficient mass ejection.

\subsubsection{Effect of accretion rate}

Figure~\ref{fig:small-structure} shows that varying the accretion rate does not dramatically modify the flow structure. Higher accretion rates raise the overall gas density scale throughout the disk and wind for both BH spin values. Figure~\ref{fig:angular} confirms this trend, showing that higher $\dot{m}$ increases the density at nearly all inclinations while leaving the angular density structure broadly similar to that of the low-$\dot{m}$ models.

The effect of accretion rate on the overall wind velocity depends somewhat on spin. For $a = 0$, higher $\dot{m}$ mildly suppresses wind velocities at low to intermediate inclinations. For $a = 0.9$, the polar wind velocity remains high because the jet and magnetized outflow dominate the low-inclination dynamics, unaffected by the increased density. In both cases, however, the velocity changes are modest compared with the dramatic increase in wind density.

The wind loss rate scales approximately as
\begin{equation}
    \frac{\mathrm{d}\dot{M}_{\rm wind}}{\mathrm{d}\Omega} \propto \rho_{\rm wind} \, r^2 \, v^r.
\end{equation}
Between the low-$\dot{m}$ and high-$\dot{m}$ fiducial models, the BH accretion rate increases by a factor of $\sim25$, while the wind density increases by a factor of $\sim100$. The outflow velocity changes only mildly. This explains the lower accretion ratio at high mass-supply rates (Section~\ref{sec:accretion_ratio}): as the wind loss rate grows faster than $\dot{M}_{\rm acc}$, a larger fraction of the gas is ejected rather than accreted. At higher $\dot{m}$, the denser outflow naturally carries greater kinetic power as well (Section~\ref{sec:wind_energy}).

At very high accretion rates ($\dot{m}\gtrsim100$), the jet power increases slowly because the magnetic flux threading the horizon increase slowly (Section~\ref{sec:jet_energy}). This turnover marks the hyper-Eddington regime, where the inner accretion flow ineffectively confine additional vertical magnetic flux. Beyond this point, additional mass supply primarily drives stronger winds rather than a more powerful jet: the wind kinetic energy continues to grow, while the jet's electromagnetic contribution plateaus.

Lastly, we also calculate the radial profiles of the mass fluxes, disk gas density, inflow velocity, and specific angular momentum. We refer the reader to Appendix \ref{ap:radial_structure} for more details.

\subsection{Photospheres, isotropic luminosity, and beaming}
\label{sec:rad_beaming}

\begin{figure*}
    \centering
    \includegraphics[width=0.9\textwidth]{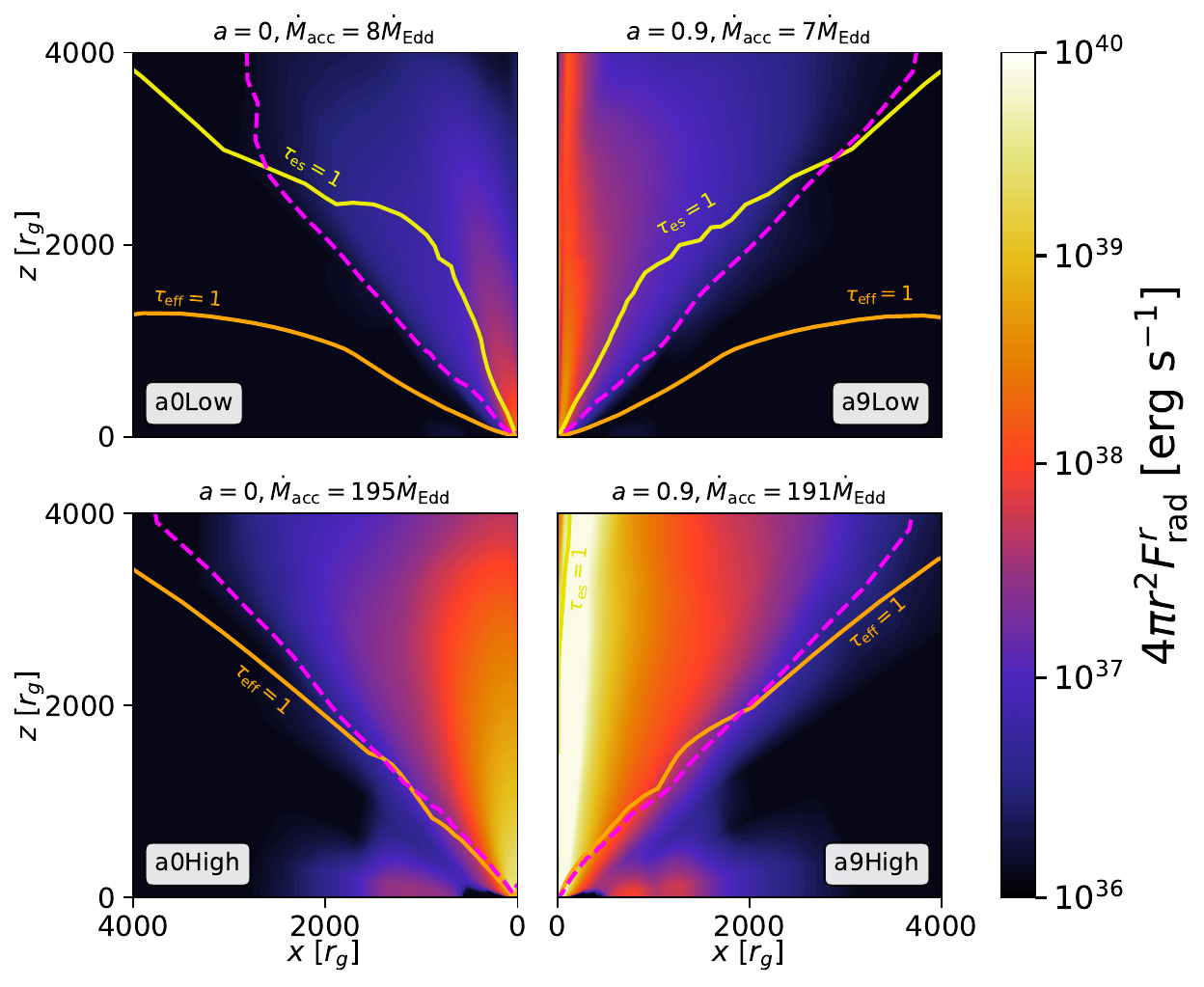}
    \caption{Time- and azimuthally-averaged isotropic-equivalent radiation luminosity $L_{\rm{rad, iso}}$ out to $r = 4,000 r_g$ for the four fiducial models. Yellow and orange contours show the electron-scattering ($\tau_{\rm{es}} = 1$) photosphere, effective photospheres ($\tau_{\rm{eff}} = 1$), respectively. Pink dashed lines show the boundary between thermally bound and unbound wind regions ($-hu_t = 1$).  }
    \label{fig:Rad-structure}
\end{figure*}

\begin{figure}
    \vspace*{-0.2cm}
	\includegraphics[width=\columnwidth]{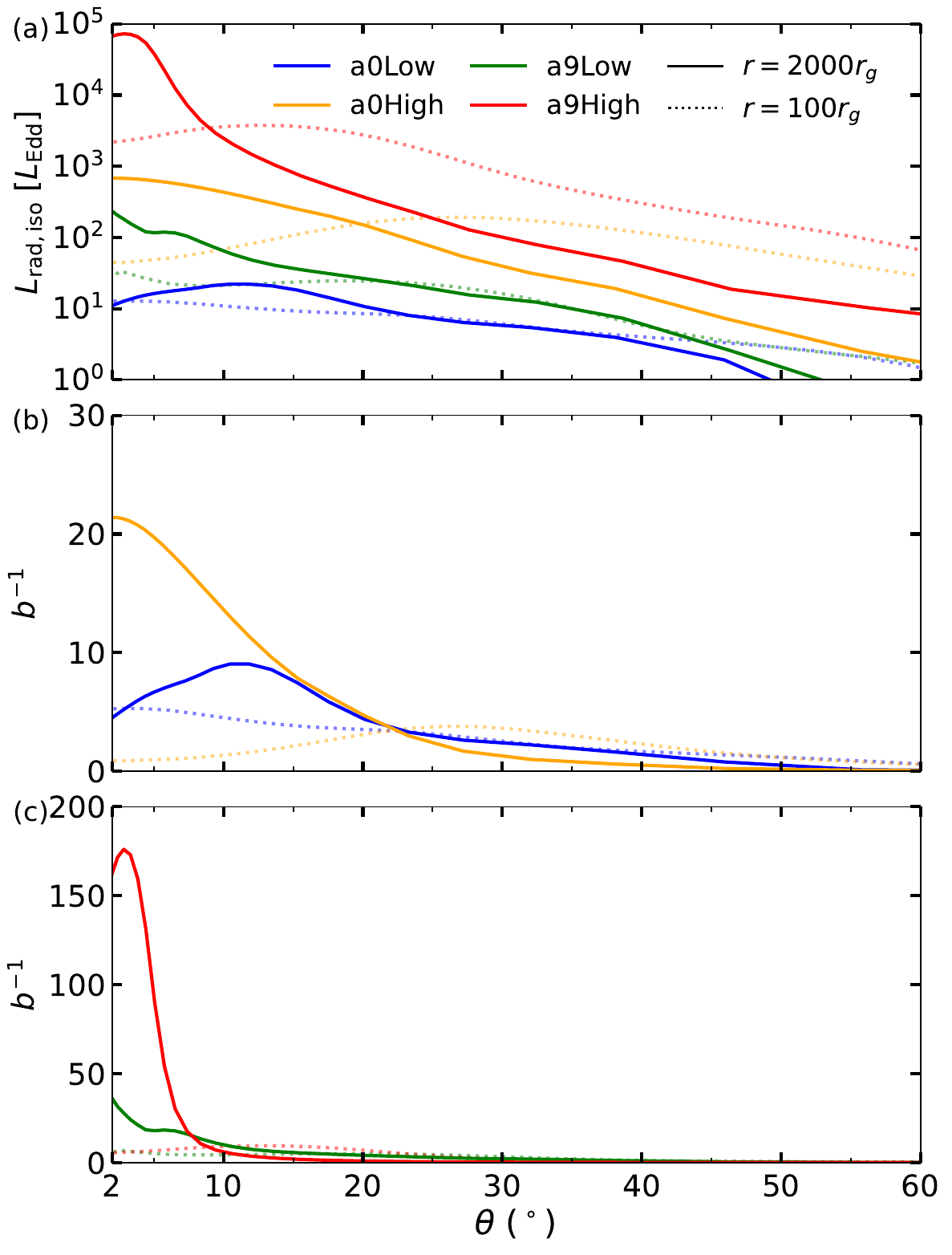}
    \vspace*{-0.2cm}
    \caption{Panels (a)–(c): quantities as a function of inclination angle $\theta$ at $r = 2000r_g$ (solid lines) and $100r_g$ (dotted lines) for the four fiducial models. (a) Isotropic-equivalent radiative luminosity $L_{\mathrm{rad,iso}}/L_{\mathrm{Edd}}$. (b) Inverse beaming factor $b^{-1}(\theta)$ for the $a = 0$ models. (c) Inverse beaming factor $b^{-1}(\theta)$ for the $a = 0.9$ models. Larger $b^{-1}$ indicates stronger beaming. }
    \vspace*{-0.2cm}
    \label{fig:angular_beaming}
\end{figure}

\subsubsection{Photosphere properties and dependence on BH parameters}
\label{sec:photosphere}

The location and shape of the photospheres determine where and how radiation escapes from the accretion flow. We define two photospheres: the electron-scattering photosphere ($\tau_{\rm es}=1$) and the effective photosphere ($\tau_{\rm eff}=1$), which accounts for both scattering and free-free absorption. Figure~\ref{fig:Rad-structure} shows the time- and $\phi$-averaged distribution of the isotropic-equivalent radiation luminosity $L_{\rm rad,iso}$ over a large part of the simulation domain, with the two photospheres overlaid for each fiducial model\footnote{Jet gas densities are subject to numerical floor contamination, so the photosphere in jet regions should be interpreted with caution.}.  Super-Eddington disks and winds are optically thick at high inclinations, while an optically thin funnel (thin for absorption, though possibly still thick for scattering) forms in the polar region.

\textbf{Dependence on BH spin:}
BH spin only moderately affects the photosphere structure by changing the polar density structure. For $a=0.9$, the relativistic jet evacuates gas from the polar region. Consequently, the electron-scattering photosphere is pushed to larger polar angles, leaving a clearer path for radiation to escape along the rotation axis. For $a=0$, no jet forms, and the polar region remains filled with slower, denser wind, resulting in a slightly narrower scattering funnel. The overall difference between the two spin states is modest, reflecting that the jet's primary effect is localized to the polar region.

\textbf{Dependence on BH accretion rate:}
The accretion rate has a stronger effect on the photosphere structure than BH spin. Increasing $\dot{m}$ raises the gas density throughout the disk and wind, pushing both the electron-scattering and effective photospheres to larger radii. In the low-$\dot{m}$ models, \texttt{a0Low} and \texttt{a9Low}, the scattering photosphere remains within $r \sim 1000-2000\,r_g$ connecting to the BH at low inclinations, allowing radiation to escape from the inner funnel. In the high-$\dot{m}$ models, \texttt{a0High} and \texttt{a9High}, the denser wind makes even the polar ultrafast outflow optically thick to electron scattering, causing the scattering photosphere to expand substantially. In the most extreme non-spinning case, model \texttt{a0High}, the polar funnel is almost completely closed by the optically thick wind.

In summary, BH spin determines whether a relativistic jet opens a low-density polar funnel that mildly affects the location of the scattering photosphere, whereas the accretion rate primarily controls the optical depth and overall size of the photospheres.

\subsubsection{Isotropic luminosity and dependence on BH parameters}

The isotropic-equivalent radiative luminosity $L_{\rm rad,iso}$, inferred from the observed flux under the assumption of isotropic emission, provides the most direct link between our simulations and ULX observations. Understanding its dependence on BH parameters and inclination is essential for interpreting the observed diversity of ULXs.

Figure~\ref{fig:angular_beaming} panel (a) shows $L_{\rm rad,iso}$ as a function of inclination angle $\theta$ at $r=2000\,r_g$ (solid lines) and $100\,r_g$ (dotted lines) for the four fiducial models. At inclination angles $\theta \lesssim 40^\circ$, all models produce $L_{\rm rad,iso} \gg L_{\rm Edd}$, with values generally increasing toward the pole.

\textbf{Dependence on radius:} The angular distribution of the radiation becomes increasingly anisotropic as it propagates outward. Within the optically thick wind, photons are advected outward with the gas while preferentially diffusing toward directions of lower optical depth. The denser wind at higher inclinations redirects radiation toward the low-density polar funnel. Once the photons cross the scattering photosphere, they free-stream outward and preserve this polar distribution. As a result, the radiation field becomes more strongly beamed toward low inclinations at large radii.

\textbf{Dependence on BH spin:} Notably, at fixed $\dot{m}$, the high-spin models produce much larger $L_{\rm rad,iso}$ at low inclinations. This is consistent with the higher radiative efficiencies found in Section~\ref{sec:luminosity}. In the $a=0.9$ models, the jets and fast wind evacuate gas from the polar region, drilling an optically thinner funnel that allows more radiation to escape toward low-inclination observers.

\textbf{Dependence on BH accretion rate:} Increasing 
accretion rate also enhances the observed luminosity. For the $a = 0.9$ models, increasing $\dot{m}$ by a factor of $\sim25$ (from \texttt{a9Low} to \texttt{a9High}) raises $L_{\rm rad,iso}$ by a factor of $\sim100$ -- $400$ in the UFO region at low inclinations ($\theta \lesssim 20^\circ$; Figure~\ref{fig:angular_beaming} panel a). This is consistent with the super-linear scaling of $L_{\rm rad}$ with $\dot{m}$ found in Section~\ref{sec:luminosity}. For the $a = 0$ models, the same increase in $\dot{m}$ produces a more modest luminosity enhancement, by a factor of $\sim10$ -- $100$ at low inclinations, reflecting the near-linear scaling in the non-spinning case.

In the high-$\dot{m}$ models, the fast polar outflow remains optically thick to electron scattering even at low inclinations, yet the observed luminosity can still be extremely large. This is because the outflow is thick to scattering but thin to absorption: photons are scattered many times but not destroyed, and gradually diffuse out through the polar funnel while being advected outward by the dense flow. For face-on observers, this produces a strongly enhanced apparent luminosity, offering a possible explanation for how hyper-Eddington sources can appear to shine far above the Eddington limit.

\begin{figure}
    \vspace*{-0.2cm}
    \includegraphics[width=\columnwidth]{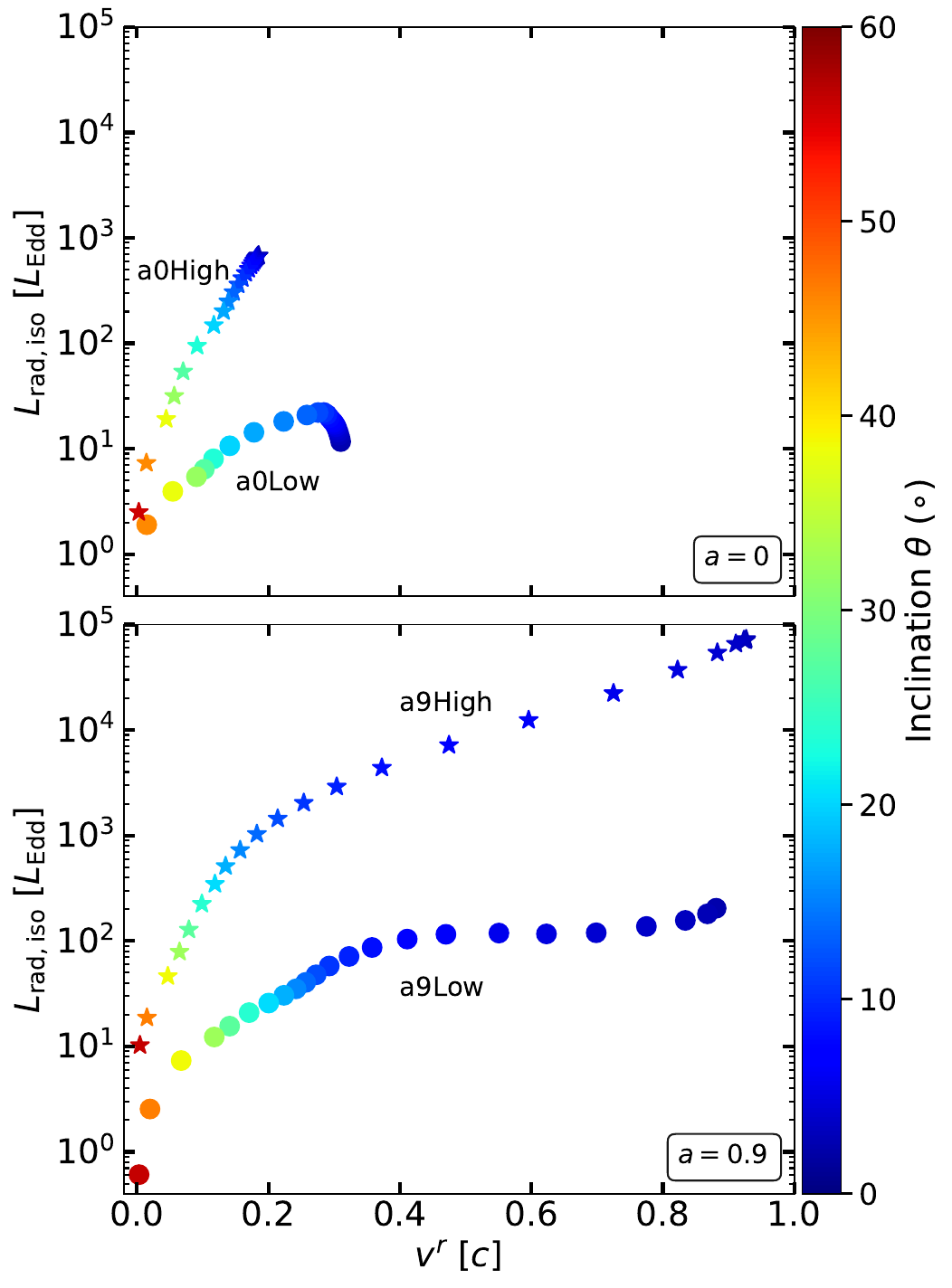}
    \vspace*{-0.4cm}
    \caption{ Isotropic-equivalent radiative luminosity $L_{\rm rad,iso}$ and radial velocity $v^r$ of the outflow at different inclination angles, measured at $r=2000 r_g$. The top panel shows the $a=0$ fiducial models, and the bottom panel shows the $a=0.9$ fiducial models. Colors indicate the inclination angles. Circles denote the low-$\dot{m}$ models, \texttt{a0Low} and \texttt{a9Low}, while stars denote the high-$\dot{m}$ models, \texttt{a0High} and \texttt{a9High}. 
     }
     \vspace*{-0.2cm}
    \label{fig:Lvr}
\end{figure}

Figure~\ref{fig:Lvr} further shows the relation between $L_{\rm rad,iso}$, the outflow radial velocity $v^r$, and inclination $\theta$ at $r=2000\, r_g$.
The apparent luminosity and outflow velocity do not vary independently. Instead, they are strongly correlated with inclination. At low inclination ($\theta \lesssim 30^\circ$), sight lines with large $L_{\rm rad,iso}$ generally pass through the faster, lower-density polar outflow. At higher inclinations, the denser and slower wind limits the escaping radiation. This trend is broadly consistent with previous simulations of super-Eddington accretion. For example, \citet{Sadowski.2015} and \citet{Ogawa.2017} reported that nearly face-on observers infer larger apparent luminosity and observe faster outflows. In our simulations, rapid BH spin and high accretion rate further enhance the apparent luminosity. The resulting coupling between inclination, outflow speed, and apparent luminosity provides a useful observational diagnostic, suggesting that very luminous sources accompanied by fast outflows (UFOs or relativistic jets) are preferentially associated with high-spin and/or high-$\dot{m}$ systems viewed close to the polar axis.

\subsubsection{Beaming of escaping radiation}

The escape of radiation from super-Eddington disks is highly anisotropic. Photons escape preferentially through the lower-density, faster polar outflow, whereas the dense, optically thick slow wind at high inclinations traps and redirects them. This anisotropy produces strong variations in the apparent luminosity with viewing angle.

We quantify the anisotropy using the beaming factor, $b(\theta) = L_{\rm rad}/L_{\rm rad,iso}(\theta)$. Figures~\ref{fig:angular_beaming} panels (b) and (c) show the inverse beaming factor $b^{-1}(\theta)$ at $r=2000\,r_g$ (solid lines) and $100\,r_g$ (dotted lines) for the $a=0$ and $a=0.9$ fiducial models, respectively. Larger $b^{-1}$ corresponds to stronger beaming.

Both high accretion rate and high BH spin strengthen beaming. A higher $\dot{m}$ increases wind density and narrows the funnel, concentrating the radiation into a smaller solid angle. In the highest-$\dot{m}$ models, \texttt{a0High} and \texttt{a9High}, the beaming is strongest, with $b^{-1}>10$ at $\theta < 10^\circ$ for $a=0$ and $b^{-1}>100$ for $a=0.9$ at $r=2000\, r_g$. High spin further enhances beaming because the jet opens a narrow, optically thin funnel along the pole.

We notice that for the \texttt{a0Low} model (blue curve in Figures~\ref{fig:angular_beaming} panels b), the peak beaming is offset from the pole by $\sim10^\circ$ at low $\dot{m}$. This offset approximately follows the opening angle of the inner scattering photosphere, although it may also be partly affected by an artifact of the M1 closure in optically thin regions.

The strong beaming implies that a stellar-mass BH in a super-Eddington MAD state viewed face-on ($\theta \lesssim 10^\circ$) can have an isotropic-equivalent luminosities exceeding $100\,L_{\rm Edd}$. Conversely, the same source viewed edge-on would appear fainter. This viewing-angle dependence provides a natural explanation for the wide range of inferred luminosities in ULXs without requiring a wide spread in BH mass. We aim to test this prediction in future studies.

\section{Discussion}
\label{sec:discussion}

\subsection{Implications for ULXs and their nebulae}

Our scaling relations enable direct comparisons between simulations of super-Eddington accretion around stellar-mass BHs and multi-wavelength observations of ULXs. Below we connect our results to three key observational diagnostics: ULX outflows, ULX nebulae, and some of the most luminous ULXs.

\subsubsection{ULX ultrafast outflows}

Many ULXs exhibit UFOs in their X-ray spectra with velocities $v \gtrsim 0.1-0.2c$ \citep[e.g.,][]{Pinto.2016, Pinto.2017, Pinto.2020}. In our simulations, both $a=0$ and $a=0.9$ models produce winds with velocities in this range at low inclinations ($\theta \lesssim 30^\circ$). However, for $a=0.9$, wind velocities reach $0.4-0.5c$ even outside the jet region (Figure~\ref{fig:angular}a). Thus, detection of UFOs with $v > 0.3c$ would favor a rapidly spinning BH interpretation, while velocities of $\sim0.1-0.2c$ alone cannot distinguish between the high and low spin states.

\subsubsection{ULX nebulae: kinetic feedback}

A growing number of ULXs are surrounded by optical bubbles inflated by mechanical feedback from their outflows \citep{Pakull.2002, Feng.2008}. These nebulae provide an indirect estimate of the time-averaged kinetic power injected into the surrounding interstellar medium, with inferred $L_{\rm kin}$ ranging from $\sim10^{39}$ to $10^{41}$ erg~s$^{-1}$ \citep[e.g.,][]{Abolmasov.2007, Walton.2016, Pinto.2016,Pinto.2017,Pinto.2020}.

Our scaling relations (Equations~\ref{eq:luminosity_a0_kin} and~\ref{eq:luminosity_a9_kin}) allow the observed $L_{\rm kin}$ to be translated into constraints on BH spin and accretion rate. 
For example, the large bubble surrounding the ULX in NGC~5585 has an inferred time-averaged kinetic power of $L_{\rm kin} \sim 2\times 10^{40}$ erg s$^{-1}$ \citep{Soria.2021}. Assuming a $10\,M_{\odot}$ BH, our scaling relations imply $\dot{m}\sim 10$ for a high-spin BH, or a substantially larger accretion rate, $\dot{m}\sim 50$, for $a=0$. A rapidly spinning BH thus provides an explanation for the powerful nebula without requiring an extreme accretion rate. By contrast, Holmberg II X-1 has an estimated nebular power of $L_{\rm kin}=0.7 \times10^{39}$ erg s$^{-1}$ \citep{Abolmasov.2007}.  For a $10\,M_{\odot}$ BH, this low value is more consistent with a BH of modest spin, accreting at a rate closer to the Eddington limit.

For a given $\dot{m}$, higher spin produces both higher radiative and kinetic output, as well as faster polar outflows. The degeneracy between BH spin and accretion rate can be reduced by combining nebular kinetic power measurements with X-ray luminosity and spectral information from the ULX itself. This approach warrants further exploration.

\subsubsection{Hyper-luminous X-ray sources}

Hyper-luminous X-ray sources (HLXs) have isotropic-equivalent luminosities exceeding $10^{41}$ erg s$^{-1}$ ($\gtrsim 80\,L_{\rm Edd}$ for a $10\,M_\odot$ BH). They can be powered by hyper-Eddington accretion onto neutron stars \citep[e.g. NGC 5907 ULX1,][]{Israel.2017}, stellar-mass or intermediate-mass BHs \citep[e.g. HLX-1,][SS433]{Farrell.2009}. For an $a=0$ BH viewed at $\theta\sim10^\circ$, an accretion rate of $\dot{m}>100$ is required to reach an isotropic-equivalent luminosity $L_{\rm rad,iso}\gtrsim100\,L_{\rm Edd}$, sufficient to explain HLX luminosities. Rapid BH spin substantially reduces this requirement by increasing both the intrinsic radiative output and the degree of polar beaming.

\subsection{Implication for black hole spin evolution in HMXBs}

Wind-fed BH high-mass X-ray binaries (BH-HMXBs) consist of a stellar-mass BH accreting from the strong stellar winds of a massive companion star. 
The systems may have undergone an earler phase of Roche-lobe overflow (RLO), during which the BH could have accreted at super-Eddington rates \citep{Moreno.2011}. The fraction of gas mass that reaches the BH presented in this work is therefore important for understanding mass growth and spin evolution of the BH.

The three dynamically confirmed wind-fed BH-HMXBs, Cygnus X-1, LMC X-1, and M33 X-7, are all inferred to host rapidly spinning BHs \citep[e.g.,][]{McClintock.2014,Reynolds.2014,Reynolds.2021}. This appears to contrast with the generally low effective BH spins observed in many gravitational-wave BH binaries \citep[e.g.,][]{Abbott.2019,Miller.2020,Roulet.2021,Abbott.2023}. Stellar evolution models predict a low natal spins for the first-born BH in binaries due to efficient angular momentum transport. The origin of the high BH spins in wind-fed BH-HMXBs therefore remains an open question.

\citet{Xing.2025} recently investigated whether an earlier phase of super-Eddington accretion could account for the high BH spins of wind-fed BH-HMXBs. They incorporated a BH accretion prescription motivated by our simulations into detailed binary evolution and population synthesis calculations. They found that a substantial fraction of wind-fed BH-HMXBs underwent RLO phase after BH formation, during which the BH could accrete substantial amount of gas. 
Under the moderately super-Eddington accretion prescription obtained from our simulations, BHs can reach spins of $a\sim 0.3-0.6$, assuming that the accreted gas carries the specific angular momentum at the innermost stable circular orbit. However, reproducing the highest inferred BH spins requires substantially more conservative accretion.

Our super-Eddington MAD simulations show that approximately $7-30\%$ of the supplied gas reaches the BH. However, the strong magnetic fields in MADs extract angular momentum from the BH \citep[e.g.][]{Lowell.2024}
A SANE disk with small amount of accumulated magnetic flux has weak magetic torque powering wind and could retain a larger fraction of the supplied gas.
Such conditions allow more efficient BH spin-up and can spin up the BH to a high equilibrium spin value \citep{Gammie.2004}, although this possibility remains to be explored with super-Eddington SANE disk simulations. 

\subsection{Context in previous super-Eddington simulations}
\label{sec:literature}

Before comparing to previous work, we briefly summarize our main results relevant to this discussion. Over the mass range explored here, $5-30\,M_\odot$, BH mass produces no identifiable effect on any of the measured quantities. For $a=0$, the radiative and kinetic efficiencies remain approximately constant over $\dot{m}=1-2000$, with $\eta_{\rm rad} \sim 1-3\%$ and $\eta_{\rm kin} \sim 2-4\%$. For $a=0.9$, both efficiencies increase with accretion rate: $\eta_{\rm rad}$ ranges from $\sim8\%$ to $\sim18\%$, while $\eta_{\rm kin}$ reaches $\sim60\%$ at the highest $\dot{m}$. Our simulations produce higher radiative and kinetic efficiencies than most of the SANE disk simulations presented in Figure \ref{fig:literature}. The BH accretes only $\sim7-30\%$ of the supplied gas, with the remainder ejected in winds. Radiation is strongly beamed along the polar funnel, with $b^{-1}$ exceeding $100$ for high-spin, high-$\dot{m}$ models viewed nearly face-on.

\subsubsection{Previous non-MAD simulations}

Previous super-Eddington simulations without magnetic fields, or with insufficient coherent magnetic flux to reach the MAD state, generally report lower wind mass-loss rates and kinetic luminosities than our MAD models. For example, \citet{Yoshioka.2022} simulated non-magnetized super-Eddington disks with $\dot{m} \sim 11-38$ and found accretion ratios $\dot{M}_{\rm acc}/\dot{M}_{\rm t} \sim 0.62-0.92$, substantially larger than the values of $\sim0.2-0.3$ found in our MAD models at similar $\dot{m}$. Their kinetic luminosities were correspondingly lower. This comparison suggests that strong, large-scale magnetic fields in the MAD fundamentally enhance wind-launching efficiency, allowing wind to remove a larger fraction of the supplied gas and converting more of the available accretion power into kinetic energy.

\citet{Jiang.2014} reported a non-MAD simulation with $\dot{m}\sim20$ and $a=0$ achieving $\eta_{\rm rad} \sim 4.5\%$, slightly higher than our $a=0$ MAD value ($\sim1-3\%$).  The authors attributed their relatively high radiative efficiency to MRI-driven magnetic turbulence, which advects radiation vertically from the disk midplane to the photosphere more rapidly than radiative diffusion. In our MAD simulations, however, strong global magnetic torques efficiently remove angular momentum from the disk \citep[see Appendix \ref{ap:disc_rotation}, see also][]{Lowell.2024}, substantially shortening the inward advection timescale. This rapid inflow may reduce the time available for local MRI-driven turbulence to transport radiation vertically to the disk surface. The difference in $\eta_{\rm rad}$ may also partly reflect our more restrictive definition of $L_{\rm rad}$, which excludes regions inside the effective photosphere, as well as differences in the transport of radiation through the optically thick outflow.

\citet{Fragile.2025} performed simulations of initially mildly super-Eddington disks around a rapidly spinning ($a=0.9$) stellar-mass BH. Their accretion flows self-regulate through strong outflows, such that the accretion rate near the horizon approaches the Eddington rate. Their disks remain Keplerian and achieve high radiative efficiencies of $\eta_{\rm rad} \lesssim 40-70\%$. The combination of a near-Eddington inner accretion rate and strong outflows, and the long orbital time associated with the Keplerian flow may allow a large fraction of the radiation to escape. These simulations therefore represent a different accretion regime from most previous super-Eddington simulations, including those presented in this study.

\subsubsection{Previous MAD simulations}

\citet{McKinney.2015} simulated a super-Eddington MAD model with $a=0.8$ and $\dot{m}=50$, finding $\eta_{\rm rad} \sim 15\%$ for radiation escaping outside the photosphere. This value is consistent with our $a=0.9$ results at similar $\dot{m}$, for which $\eta_{\rm rad} \sim 10-15\%$. The agreement supports the conclusion that high spin leads to high radiative efficiency.

\citet{Curd.2023b} simulated MADs transitioning from super-Eddington to sub-Eddington accretion for both $a=0$ and $a=0.9$. Their super-Eddington models have higher radiative efficiencies than ours, which may be partly due to their smaller measurement radius, where a larger fraction of the outflow energy remains in radiative form, as discussed previously. Nevertheless, their results are qualitatively consistent with ours: high-spin models exhibit higher jet and radiative efficiencies, and outflow properties depend strongly on BH spin. Our work extends this picture by providing scaling relations over a broad range of $\dot{m}$ and by systematically examining the effects of BH mass within the stellar-mass range.

\citet{Kaaz.2025} carried out two GRRMHD simulations of highly magnetized, nearly MAD-state accretion disks around a $1.3\times 10^7 M_{\odot}$ SMBH, with spins $a=0$ and $0.9375$. Their initial accretion flows were remapped from a cosmological radiation-MHD simulation performed with \texttt{GIZMO}, in which the disk formed self-consistently within its galactic environment \citep{Hopkins.2024}. After radiation transport was activated, the accretion rates were sustained at $\approx 5 \dot{M}_{\rm Edd}$. Despite the very different setup, they reported very large mechanical and radiative efficiencies, with the radiative efficiency approaching $100\%$ in the $a=0.9375$ model.

\section{Summary}
\label{sec:summary}

We have presented 30 three-dimensional GRRMHD simulations of super-Eddington magnetically arrested disks around stellar-mass black holes, systematically varying BH mass ($5$, $15$, and $30\,M_{\odot}$), spin ($a=0$ and $0.9$), and accretion rate ($\dot{M}_{\rm acc} \approx 1-2000\,\dot{M}_{\rm Edd}$). Our key findings are as follows:

\begin{enumerate}
    \item \textbf{BH mass has no effect.} Over the range $5-30\,M_{\odot}$, BH mass does not systematically influence the flow structure, energy output, or mass ejection efficiency. This holds for both spin values and across all accretion rates. 

    \item \textbf{Wind-driven energy output}: In the wind, kinetic energy dominates across all models at large radii (thousands of $r_g$). Spin and accretion rate jointly regulate wind energetics, but in qualitatively different ways.
    \begin{itemize}
        \item For $a=0$, the wind kinetic, radiative, and EM efficiencies are approximately constant across $\dot{m}$: $\eta_{\rm kin} \sim 2-4\%$, $\eta_{\rm rad} \sim 1-3\%$, and $\eta_{\rm EM}$ is negligible. All luminosities scale nearly linearly with $\dot{m}$.
        \item For $a = 0.9$, the wind kinetic, radiative, and EM efficiencies increase with $\dot{m}$: $\eta_{\rm kin}$ reaches $\sim60\%$ at the highest accretion rates, $\eta_{\rm rad}$ ranges from $\sim8\%$ to $\sim18\%$, and $\eta_{\rm EM}$ grows to become comparable to $\eta_{\rm rad}$ at high $\dot{m}$. The wind luminosities scale super-linearly ($L_{\rm kin} \propto \dot{m}^{1.25}$, $L_{\rm rad} \propto \dot{m}^{1.13}$, $L_{\rm EM} \propto \dot{m}^{1.26}$). 
        \item High accretion rate alone is insufficient to achieve very high wind efficiency. High BH spin is required to make wind power grow faster than the accretion rate.
    \end{itemize}

    \item \textbf{Jet energy output: relativistic jets provide an additional energy channel in high-spin systems.} For the $a=0.9$ models, the jet power scales as $P_{\rm jet} \propto \dot{m}^{1.29}$ for $1<\dot{m}\lesssim100$, and saturates at higher accretion rates. This saturation is regulated by the magnetic flux threading the BH horizon, which peaks around $\dot{m}\sim100$ and declines at hyper-Eddington rates. The jet total efficiency $\eta_{\rm jet}$ correspondingly turns over and decreases at the highest accretion rates.

    \item \textbf{Why kinetic energy dominates the outflow?} As the outflow propagates outward, electromagnetic and radiative energy are progressively converted into kinetic energy. This largely explains why our results differ from most previous super-Eddington simulations, which typically reported radiation-dominated outflows. We evaluate the wind energetics at large radii ($r = 2000r_g$) outside the effective photosphere, where the conversion from EM and radiative to kinetic energy has largely completed, while many previous studies used smaller simulation domains and measured the energy at smaller radii. Additionally, our simulations reach the MAD state, where strong magnetic fields efficiently convert energy into kinetic form.
    
    \item \textbf{The BH accretes only a fraction of the supplied gas.} By making a robust measurement of the wind rate at large radii ($2000 r_g$), we estimate accretion ratios $\dot{M}_{\rm acc}/\dot{M}_{\rm t}$ of $\sim7\%$ to $30\%$, with the remainder ejected in winds. The ratio decreases with increasing mass supply rate and is systematically lower for $a=0.9$ than for $a=0$. We provide spin-dependent power-law fits (see Equation~\ref{eq:mass_ratio_fits}). However, additional wind launching from the outer disk may further reduce the fraction of gas ultimately reaching the BH.

    \item \textbf{Radiation is highly beamed along the polar funnel.} The beaming factor $b^{-1}$ exceeds $10$ at $\theta<10^\circ$ for $a=0$ and exceeds $100$ for $a=0.9$ at large radii in the highest-$\dot{m}$ models. This implies that a stellar-mass BH viewed face-on can can have an isotropic-equivalent luminosities exceeding $100\,L_{\rm Edd}$.

    \item \textbf{Observational predictions.} Sufficiently high accretion rates can produce HLX-level luminosities for $>10^{41}$ erg s$^{-1}$ and strong nebular kinetic power for both non-spinning and rapidly spinning BHs. Rapid BH spin substantially lowers the accretion rate required to produce powerful nebular power and extreme isotropic-equivalent luminosities. Moreover, UFOs with observed velocities $>0.3c$ would strongly suggest high BH spin. Joint constraints from $L_{\rm rad,iso}$, wind kinetic power, outflow velocity, and viewing angle provide a more reliable diagnostic.

\end{enumerate}

Our results arrive at a systematic picture of super-Eddington MADs. In the absence of BH spin, the magnetic field drives a wind but cannot launch a Blandford-Znajek jet; the wind is powered by both magnetic pressure and radiation pressure in the disk. When a fast BH spin is present, the magnetic field extracts BH rotational energy, launching a relativistic jet. Across all models, the wind carries most of the ejected mass and kinetic energy. High spin greatly enhances both the radiative output and the wind kinetic power. High Eddington ratio scales up all wind energy output and mass loss; for the jet, however, the power grows with $\dot{m}$ until the magnetic flux saturates, after which the system enters a hyper-Eddington regime where additional mass supply primarily drives stronger winds rather than a more powerful jet.

Several caveats should be noted. The turbulence in the outer disk has not been fully developed to contribute to wind launching. By focusing only on the thermally unbound component of the wind launched from the inner disk, our wind mass and energy estimates are lower limits, and time-averaging may underestimate the true energy carried by turbulent outflows. Lagrangian tracer methods could provide more accurate wind properties \citep{Yuan.2015, Yang.2023}. The M1 closure for radiation transport may affect the precise beaming profile in optically thin polar regions \citep{Asahina.2020, Asahina.2022}, though total $L_{\rm rad}$ is less affected. The role of inflow angular momentum also remains to be explored, as low-angular-momentum flows may behave differently from the Keplerian disks assumed here \citep{Kwan.2023, Galishnikova.2025}.

Despite these limitations, the physical picture that emerges is clear: BH spin is the primary driver for the efficiency of super-Eddington MADs, while accretion rate determines the overall energy scale. These two parameters jointly regulate the energy output in different ways. Higher BH spin enhances the energy efficiency, whereas increasing the accretion rate amplifies the total power and, in high-spin systems, drives efficiency to grow with the accretion rate. Consequently, rapidly spinning BHs can produce substantially more powerful outflows and stronger radiative emission. Our scaling relations provide a framework for constraining BH spin and accretion rate from multi-wavelength observations of super-Eddington sources, including ULXs. As next-generation observatories unveil the diversity of super-Eddington sources, our framework may therefore help to interpret their physical nature and clarify the role of BH spin in determining the energetics and feedback of super-Eddington accretion flows.

\begin{acknowledgments}
We thank Hua Feng, Ken Ohsuga, Enrico Ramirez-Ruiz, Alexander Tchekhovskoy, and the referee for fruitful discussions and comments. We also thank Konstantinos Kovlakas, Devina Misra and Simone S. Bavera for useful discussions at the early stages of the project.
TMK, LD, and CKK acknowledge the support from the Hong Kong Research Grants Council (HKU 17305523, N\_HKU782/23). ZX and TF acknowledge support from the Swiss National Science Foundation (project number CRSII5\_213497). MM acknowledges support via STFC grant ST/Y001699/1. FY is supported by the NSF of China (grants 12192223, 12133008, and 12361161601).
The simulations were performed in part at the University of Hong Kong on the BLACKBODY and HPC2021 computing clusters and at the University of Geneva on the Yggdrasil computing cluster. 
OpenAI GPT 5.6 was used to improve the clarity of the manuscript.
\end{acknowledgments}


\appendix

\makeatletter
\@addtoreset{equation}{section}
\@addtoreset{figure}{section}
\@addtoreset{table}{section}
\makeatother

\renewcommand{\theequation}{\Alph{section}\arabic{equation}}
\renewcommand{\thefigure}{\Alph{section}\arabic{figure}}
\renewcommand{\thetable}{\Alph{section}\arabic{table}}

\twocolumngrid

\section{More on simulation setup}
\label{ap:setup}

\begin{figure*}
    \centering
	\includegraphics[width=0.85\textwidth]{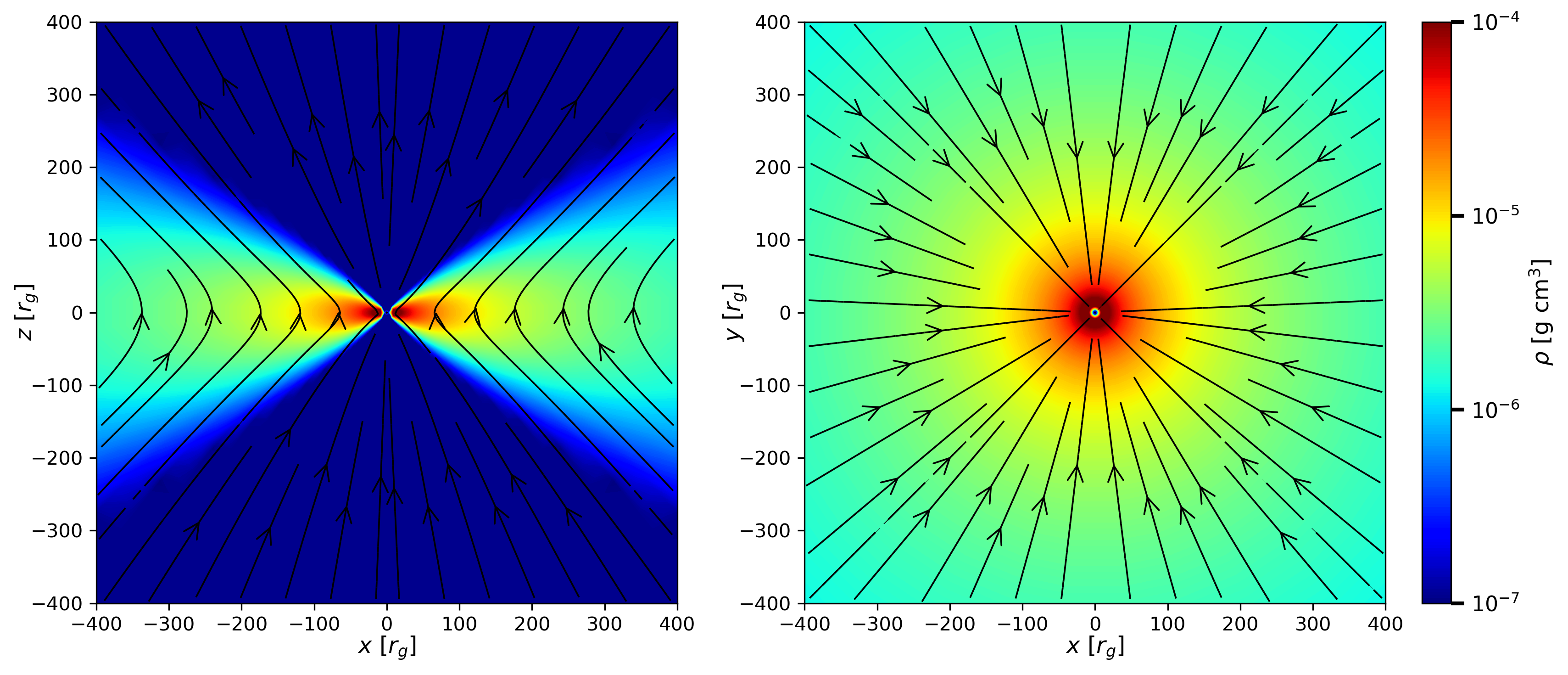}
    \vspace*{-0.2cm}
    \caption{ Initial rest-mass density $\rho$ (colour map) and magnetic field configuration (black lines). }
    \vspace*{-0.4cm}
    \label{fig:initial}
\end{figure*}

\begin{figure*}
    \centering
	\includegraphics[width=0.88\textwidth]{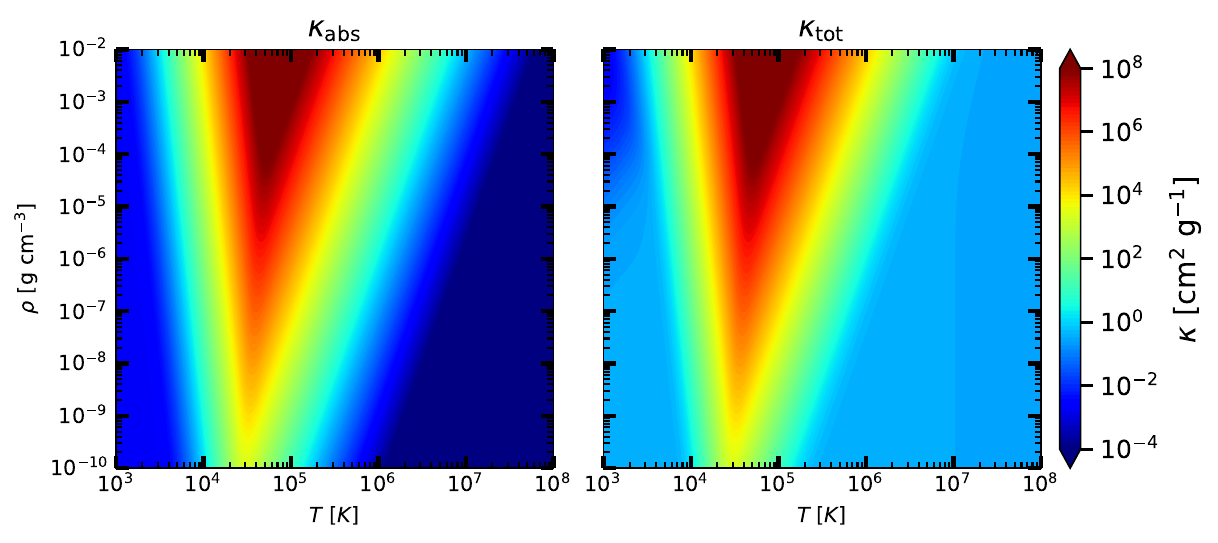}
    \vspace*{-0.4cm}
    \caption{ The absorption and total opacities at different rest-mass densities ($\rho$) and temperatures ($T=T_g=T_r$) in the simulations. }
    \label{fig:opacity}
\end{figure*}

All models use the consistent disk and magnetic field initial setup. We plot the initial gas density of the accretion flow of a model, together with magnetic field lines, in Figure \ref{fig:initial}.

The simulations include bound-free, free-free, molecular, H$^-$, and Chianti opacities as described in \citet{McKinney.2015}. The electron scattering opacity includes Klein-Nishina corrections and corrections for degenerate matter \citep{Buchler.1976, Paczynski.1983}:
\begin{equation}
    \kappa_{\rm es} = 0.2(1+X)\left(1+2.7\times10^{11}\rho T^{-2}\right)^{-1}
    \left[1+\left(\frac{T}{4.5\times10^8}\right)^{0.86}\right]^{-1},
\end{equation}
where $X=0.7$ is the hydrogen mass fraction. The gas and radiation temperatures are evolved separately, following the prescription of \citet{Sadowski.2015c}, in which the Compton energy exchange rate between electrons and photons is computed self-consistently. Figure~\ref{fig:opacity} shows the absorption and total opacities as functions of gas density and temperature, assuming $T_g = T_r$ for the purpose of this map.

We note that the free-free and bound-free absorption prescriptions adopted from \citet{McKinney.2015} use coefficients corresponding to Rosseland-mean opacities, whereas the corresponding Planck-mean absorption opacities are larger by roughly a factor of $\sim30$. Our absorption opacity may therefore be underestimated. This primarily affects the thermal coupling between gas and radiation and the location of the effective photosphere. A larger absorption opacity would move the effective photosphere outward. The effect on the overall dynamics is expected to be less important in regions where electron scattering dominates the total opacity and where the magnetic stress is the dominant outflow driving force.

To prevent vacuum violations in the jet region, we impose numerical ceilings on $b^2/\rho$, $b^2/u_g$, and $u_g/\rho$, and distance-dependent floors on $\rho$ and $u_g$. At $r>5000\,r_g$, an outer relaxation zone applies stricter ceilings and suppresses inward radial motions to prevent boundary reflection. Results beyond this radius are not used in our analysis.

\begin{figure*}
	\includegraphics[width=\textwidth]{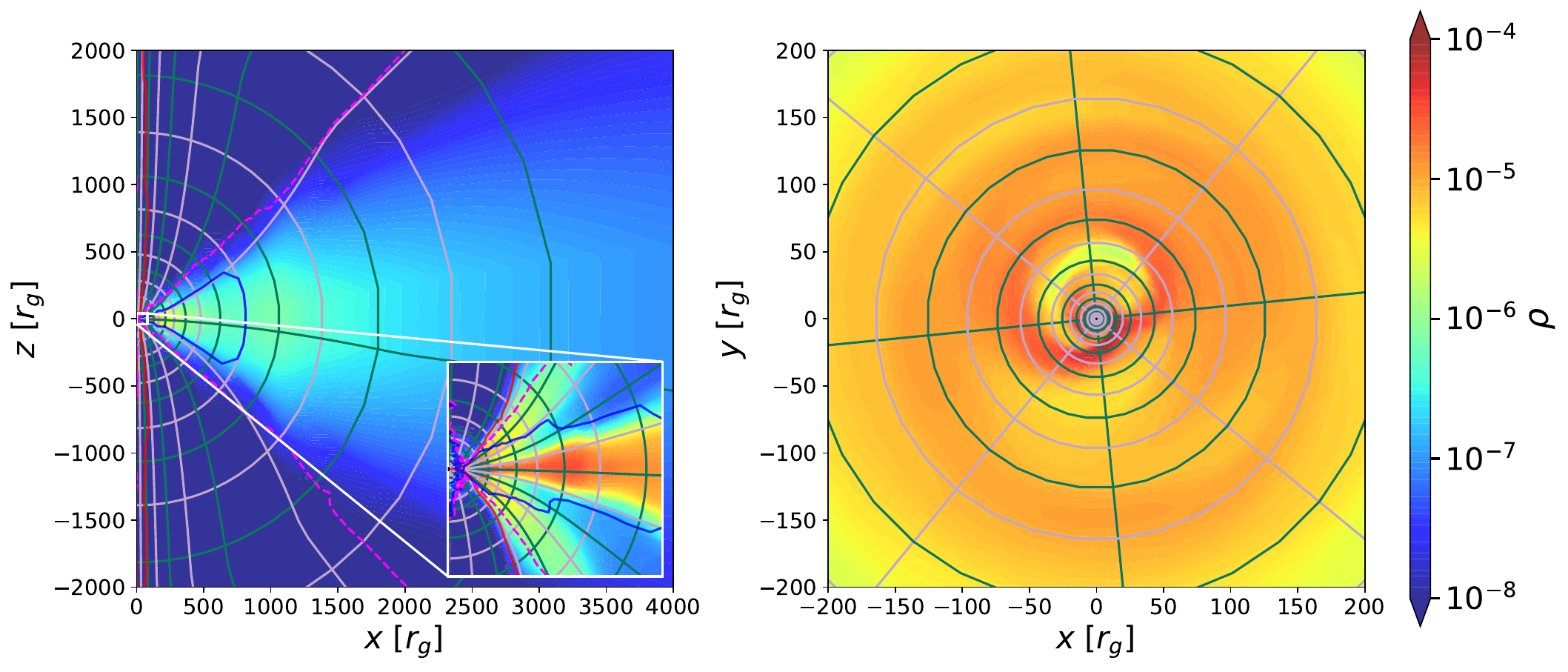}
    \vspace*{-0.6cm}
    \caption{ Representative numerical grid for an $a=0.9$ simulation with resolution $N_r \times N_{\theta} \times N_{\phi} = 128\times64\times32$, overlaid on the rest-mass density snapshot of model \texttt{m15a9r1} (\texttt{a9Low}, also see Figure \ref{fig:snapshot}). The left panel shows the x-z plane out to $2000\, r_g$, with an inset showing the inner $40\, r_g$, while the right panel shows the x-y plane out to $200\, r_g$. The red contour mark the jet-wind boundary ($\sigma=1$), the pink dashed contour mark the thermally-bound-unbound boundary ($-hu_t=1$), and the blue contour marks the inflow-outflow boundary ($v^r=0$). For clarity, only every fourth grid line is shown, alternating between dark-green and light-purple.}
    \label{fig:grid}
\end{figure*}

To illustrate the numerical grid used in our simulations, Figure \ref{fig:grid} shows the grid for model \texttt{m15a9r1}, overlaid on a snapshot at $t=40000\,t_g$. The coordinate mapping follows that of \cite{McKinney.2015}, which in turn adopts the grid-mapping prescription and boundary conditions of \cite{McKinney.2012}.

\section{MRI Quality Factor}
\label{ap:mri}

\begin{figure}
	\includegraphics[width=\columnwidth]{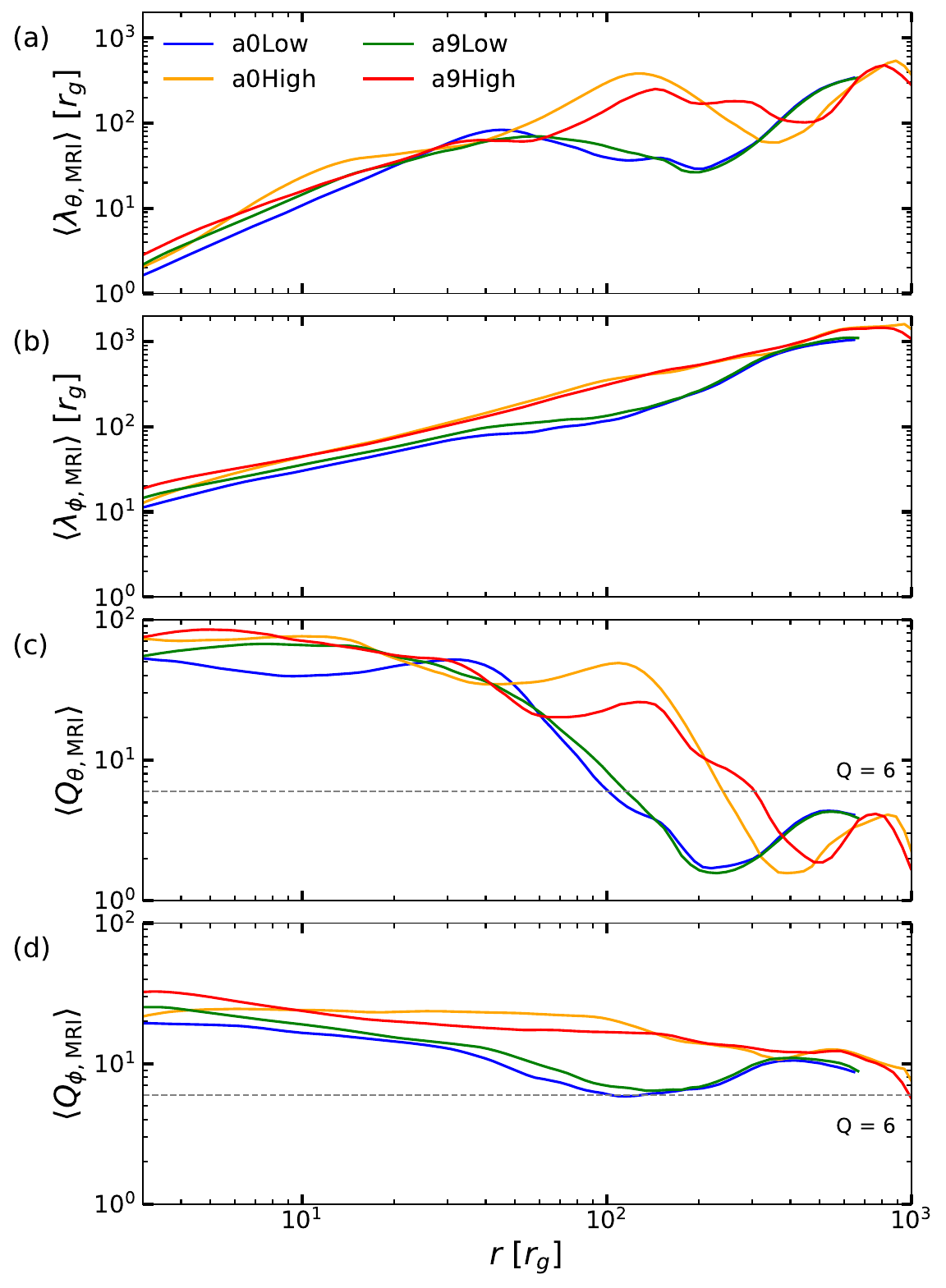}
    \vspace*{-0.6cm}
    \caption{Time- and angle-averaged radial profiles of the fastest-growing MRI wavelength in the (a) $\theta$-direction, $\lambda_{\rm \theta, MRI}$, and (b) $\phi$-direction, $\lambda_{\rm \phi, MRI}$, together with the corresponding MRI quality factor in the (c) $\theta$-direction, $Q_{\rm \theta, MRI}$, and (d) $\phi$-direction, $Q_{\rm \phi, MRI}$. The horizontal dashed lines panels (c) and (d) mark $Q_{\rm x,MRI} = 6$, which are the minimum criterion for resolving the fastest-growing MRI mode.}
    \label{fig:mri}
\end{figure}

The magnetorotational instability (MRI) drives turbulence for angular momentum transport in ionized accretion flows \citep{Balbus.1991, Balbus.1998}. To assess whether the fastest-growing MRI modes are fully resolved in our simulations, we estimate the fastest-growing wavelengths in the $\theta$- and $\phi$- directions as
\begin{equation} \label{MRI}
    \lambda_{x, \rm MRI} \approx 2\pi \frac{\langle v_{x,\rm A}\rangle_{\theta,\phi}}{\langle\Omega_r\rangle_{\theta,\phi}}
\end{equation}
with $x=\theta$ or $\phi$ \citep{Hawley.1995}. Here, $v_{x, \rm A} = \sqrt{b_x b^x/(b^2+\rho+u_g+p_g)}$ is the Alfv\'en speed in the \textit{x}-direction in Kerr-Schild coordinates, and $\Omega_r=\sqrt{v_\theta^2+v_\phi^2}/r$ is the characteristic angular frequency of the rotating flow. We calculate the quantities at each radius and time using a $\sqrt{b^2\rho}$ weighting, which emphasizes the magnetized, mass-carrying gas in the disk. Figure \ref{fig:mri} (a) and (b) show the radial profiles of $\lambda_{\rm \theta, MRI}$ and $\lambda_{\rm \phi, MRI}$, respectively.

We quantify the numerical resolution of these modes using the MRI quality factor, 
\begin{equation} \label{Q_MRI}
    Q_{x, \rm MRI} = \frac{\lambda_{x, \rm MRI}}{\Delta_{x}}
\end{equation}
where $\Delta_{x}$ is the grid cell size in the corresponding direction. The quality factor measures the number of grid cells spanning the gastest-growing MRI wavelength.
We adopt $Q_{x, \rm MRI} \geq 6$ as the criterion for resolving the MRI \citep{Sano.2004, Noble.2010, Hawley.2011}. 
Figure~\ref{fig:mri} (c) and (d) show $Q_{\rm \theta, MRI}$ and $Q_{\rm \phi, MRI}$, respectively. The values at $r=20 r_g$ and $80 r_g$ are listed for all models in Table~\ref{tab:parameters}.

The fastest-growing MRI modes are generally well resolved in both angular directions throughout the inner disk, particularly within $r\lesssim100r_g$, where the accretion flow has reached inflow equilibrium and most of our analysis of the disk structure is performed.


\section{Radial Structure of the Accretion Flow and Energy Fluxes}
\label{ap:radial_structure}

To complement the energy conversion in the outflow discussed in Section \ref{sec:energy_flux} and angular structure discussed in Section \ref{sec:structure}, we examine the radial structure of the accretion flow in the four fiducial models, focusing on the mass flow, gas density, inflow velocity, specific angular momentum, and outflow energy fluxes.

\begin{figure}
	\includegraphics[width=\columnwidth]{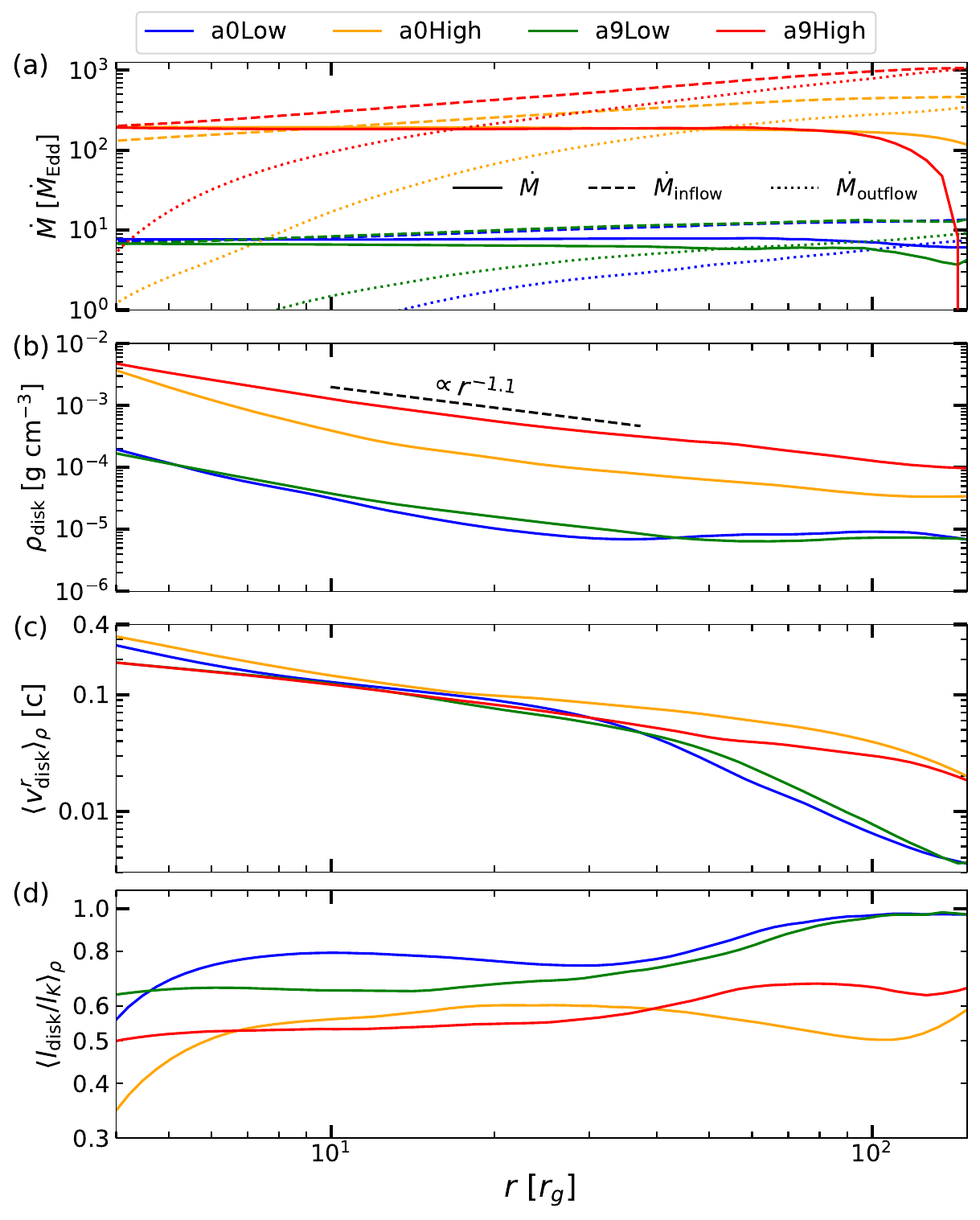}
    \vspace*{-0.6cm}
    \caption{ The time-angle-averaged radial profiles of (a) inflow (dashed lines), outflow (dotted lines), net mass flow rate (solid lines); (b) disk gas mass density $\rho_{\rm disk}$, (c) density-weighted radial velocity in the disk inflow $v^r_{\rm disk}$, and (d) density-weighted specific angular momentum in the inflow $l_{\rm disk}$, normalized by the relativistic Keplerian angular specific momentum $l_K$.}
    \label{fig:disk_radial}
\end{figure}

\subsection{Mass flow rate}

Figure \ref{fig:disk_radial}(a) shows the radial profile of the net mass accretion rate, $\dot{M}$, together with its inward and outward components, $\dot{M}_{\rm inflow}$ and $\dot{M}_{\rm outflow}$. The net accretion rate remains approximately constant within $r \lesssim 100 r_g$, indicating that the inner accretion flow has reached inflow equilibrium. At larger radii, the outward mass flux becomes increasingly important, consistent with strong mass loss from the super-Eddington disk.

Within the innermost region, $r \lesssim 10 r_g$, the outward mass flux is relatively small and most of the gas continues to accrete toward the BH. Outflow emerging from these inner radii contributes to the hot, fast polar component seen at larger radii. 
The innermost radius from which a substantial outflow is launched, $r_{\rm wind,in}$, depends on both BH spin and accretion rate. In the $a = 0.9$ and high-$\dot{m}$ models, $r_{\rm wind,in}$ shifts inwards, suggesting that stronger magnetic stresses and radiation pressure allow gas to be launched from deeper in the gravitational potential.

\subsection{Gas density} 
Figure \ref{fig:disk_radial}(b) shows the disk gas density $\rho_{\rm{disk}}$. Higher accretion rates produce systematically denser disks, whereas BH spin has comparatively little effect on the density profile at fixed accretion rate.
Across the four fiducial models, the density approximately follows a power-law profile,
\begin{equation}
    \rho_{\rm disk} \propto r^{-n},
\end{equation}
with $n \sim 1.1 - 1.3$. The similarity of the slopes among the models indicates that increasing the accretion rate primarily changes the normalization of the disk density rather than its radial structure.

\subsection{Disk inflow and rotation}
\label{ap:disc_rotation}

Figure \ref{fig:disk_radial}(c) shows the density-weighted radial velocity of the inflow disk gas, $\langle v^r_{\rm disk}\rangle_\rho$. The inflow accelerates towards the BH in all four fiducial models as angular momentum is removed from the gas.

Figure \ref{fig:disk_radial}(d) compares the density-weighted specific angular momentum of the inflowing gas, $\langle l_{\rm disk}\rangle_\rho$, with the relativistic Keplerian value $l_K$ \citep{Bardeen.1972}, 
\begin{equation}
\label{eq:l_k}
    l_K(r) = \frac{\sqrt{r} (r^2-2a\sqrt{r}+a^2)}{r \sqrt{r^2-3r+2a\sqrt{r}}} .
\end{equation}
Close to the BH, $r\lesssim50r_g$, all four fiducial models are strongly sub-Keplerian. 
In the MAD state, differential rotation of the accretion flow winds the poloidal magnetic field into a strong toroidal component, producing large Maxwell stresses that efficiently transport angular momentum outward into the winds. For the $a=0.9$ models, frame dragging further affects the innermost flow rotation and magnetic field structure close to the horizon. These magnetic stresses, together with pressure support, reduce the rotational support required for accretion and lead to strongly sub-Keplerian inner disks.

At $r > 50 r_g$, the low-$\dot{m}$ models gradually approach Keplerian rotation, whereas the high-$\dot{m}$ models remain sub-Keplerian. This indicates that rotational support becomes increasingly important in the outer disk at moderate accretion rates, while pressure support remain dynamically important over a broader radial range at high accretion rates.

\section{outflow launching}
\label{ap:launching}

\begin{figure*}
    \centering
    \includegraphics[width=0.9\textwidth]{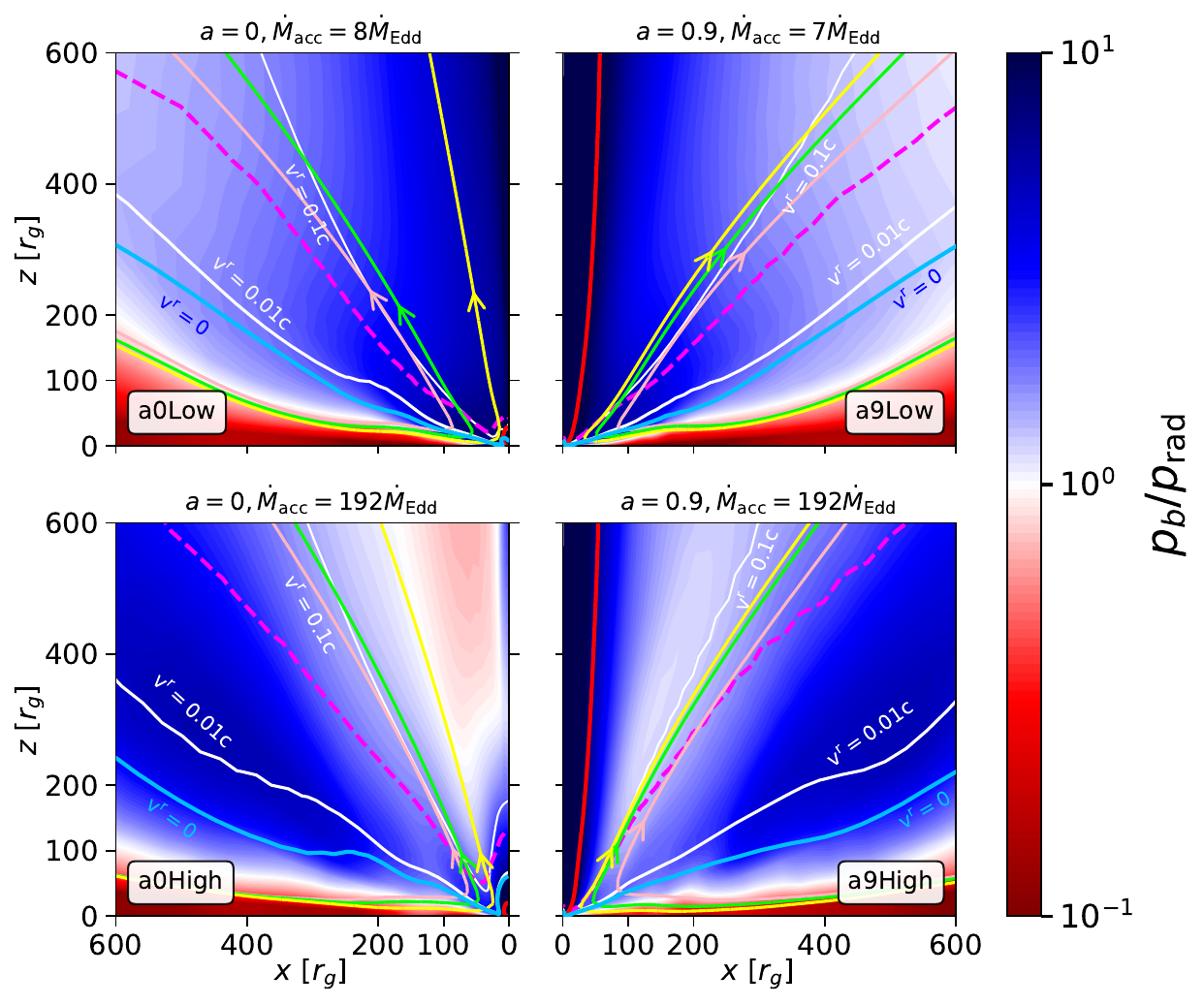}
    \vspace*{-0.2cm}
    \caption{Time-$\phi$-averaged maps of the ratio of magnetic to radiation pressure, $p_b/p_{\rm rad}$, for the four fiducial models. 
    Red regions indicate $p_{\rm rad} > p_b$, blue regions indicate $p_b>p_{\rm rad}$, and white regions correspond to approximate equipartition. Magnetic pressure dominates along the disk surface, throughout much of the inner wind, and within the relativistic jet. Colored curves show streamlines of the density-weighted, time-averaged poloidal velocity field, $\langle v \rangle_\rho$ seeded near the disk surface at $r \sim 30$ (yellow), $50$ (green), and $100$ (pink). The sky-blue contour marks the inflow-outflow boundary, $v^r=0$, while white contours show selected values of $v^r$.
    }
    \label{fig:pressure_structure}
\end{figure*}

Figure \ref{fig:pressure_structure} presents time-$\phi$-averaged maps of the ratio of magnetic to radiation pressure, $p_b/p_{\rm rad}$, for the four fiducial models.
The gas thermal pressure is generally subdominant compared with both magnetic and radiation pressure.

Magnetic pressure dominates over radiation pressure along the disk surface, throughout much of the inner wind, and within the relativistic jet when present.
In these regions, $p_b$ can exceed $p_{\rm rad}$ by up to approximately an order of magnitude. By contrast, radiation pressure dominates primarily within the dense equatorial inflow. The dominance of magnetic pressure along the disk surface indicates that magnetic forces (stress) play an important role in lifting and accelerating gas from the disk. 
In the low-$\dot{m}$ models, \texttt{a0Low} and \texttt{a9Low}, the outflow remains predominantly magnetically dominated throughout most of the inner $r<600 r_g$ region.

At high accretion rates, radiation pressure becomes increasingly important in the polar outflow and can become comparable to, or locally exceed, the magnetic pressure. In model \texttt{a0High}, the optically thick wind fills the funnel and traps radiation, maintaining a large $p_{\rm rad}$ at low inclinations. 
As a result, $p_b \lesssim p_{\rm rad}$ over much of the polar outflow beyond a few hundred $r_g$. By contrast, the $a=0.9$ models retain a wide, magnetically dominated spine along the rotational axis even at high $\dot{m}$, owing to the strong magnetic field associated with the relativistic jet.

To illustrate the geometry of the outflow, Figure \ref{fig:pressure_structure} also shows streamlines of the density-weighted, time-averaged poloidal velocity field,
\begin{equation}
    \langle v \rangle_\rho = (\langle v^r \rangle_\rho, \langle v^\theta \rangle_\rho),
\end{equation}
seeded near the disk surface at $r \simeq 30$, $50$, and $100 r_g$.
Here $v^\theta \equiv \sqrt{g_{\theta\theta}} u^\theta/u^t$. These streamlines represent the mean flow pattern rather than the actual trajectories of individual gas elements.
Near the equatorial plane, the gas flows inward toward the BH through MRI turbulence. At the disk surface, the streamlines cross the inflow-outflow boundary and bend toward the polar region, illustrating how gas is lifted from the disk and incorporated into the wind.

The disk outside $r > 100 r_g$ has not fully reached inflow equilibrium, so streamlines originating from these larger radii should be interpreted cautiously when assessing wind launching in quasi-steady state. By contrast, streamlines emerging from $r \lesssim 100 r_g$ originate within the quasi-steady inner accretion flow and connect to fast outflowing regions at larger radii, enabling robust measurements of wind fluxes far from the BH. Overall, the pressure structure and flow geometry are consistent with magnetic stresses playing the dominant role in launching the wind from the disk surface, while radiation pressure becomes increasingly important in accelerating the polar outflow at high accretion rates.

\begin{figure*}
    \centering
    \includegraphics[width=0.82\textwidth]{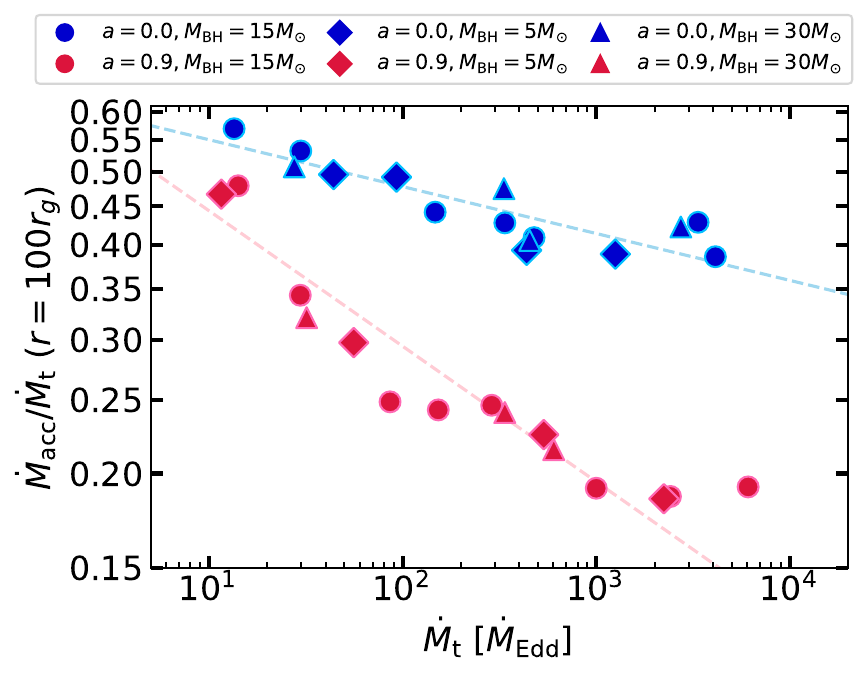}
    \vspace*{-0.4cm}
    \caption{Ratio of BH accretion rate to total mass-supply rate, $\dot{M}_{\rm acc}/\dot{M}_{\rm t}$, as a function of $\dot{M}_{\rm t}$. The total mass-supply is defined as $\dot{M}_{\rm t}=\dot{M}_{\rm acc}+\dot{M}_{\rm wind}$, where the wind rate is measured for all wind at $r = 100 r_g$.   }
    \label{fig:accretion_ratio_110}
\end{figure*}

Figure~\ref{fig:accretion_ratio_110} shows the accretion ratio $\dot{M}_{\rm acc}/\dot{M}_{t}$ where $\dot{M}_{\rm wind}$ is measured for all outflowing, quasi-steady winds at $r=100\,r_g$. Because substantially additional winds can launched at larger radii, here $\dot{M}_{t}$ is a lower limit, and therefore the accretion ratio $\dot{M}_{\rm acc}/\dot{M}_{t}$ is an upper limit.
The BH accretes around $15\%$ to $60\%$ when winds launched at larger radii are not included. Over the range $\dot{M}_{t}/\dot{M}_{\rm Edd}=10$ to $10^4$, the accretion ratio decreases from $\sim60\%$ to $\sim40\%$ for $a=0$, and from $\sim 50\%$ to $\sim15\%$ for $a=0.9$.
We fit the following power-law relations for $10 \lesssim \dot{M}_{t}/\dot{M}_{\rm Edd} \lesssim 10^4$:
\begin{align}
    \dot{M}_{\rm acc}/\dot{M}_{t} \ (r=100\, r_g) &= 
    \begin{cases}
        0.64\,\bigl(\dot{M}_{t}/\dot{M}_{\rm Edd}\bigr)^{-0.06}, & a = 0, \\[0.2cm]
        0.67\,\bigl(\dot{M}_{t}/\dot{M}_{\rm Edd}\bigr)^{-0.18}, & a = 0.9.
    \end{cases}
    \label{eq:mass_ratio_fits_100}
\end{align}

\section{Luminosity and Efficiency Tables}
\label{ap:table}

Tables~\ref{tab:parameters} and~\ref{tab:jet_parameters} list the time-averaged luminosities and efficiencies of the thermally unbound wind for all models and those of the relativistic jet for the $a=0.9$ model, respectively. All wind and jet luminosities and efficiencies are measured at $r=2000\,r_g$. We distinguish the radiative luminosity evaluated outside the effective photosphere for free-free only absorption, $L_{\rm rad,ff}\equiv L_{\rm rad}(\tau_{\rm eff,ff}<1)$ (shown in Figure \ref{fig:luminosity}), from that evaluated outside the effective photosphere accounting for free-free and bound-free absorption, $L_{\rm rad,abs}\equiv L_{\rm rad}(\tau_{\rm eff,abs}<1)$. 
The difference between $L_{\rm rad,ff}$ and $L_{\rm rad,abs}$ mainly reflects the different size and geometry of the corresponding effective photospheres. This effect is strongest in the highest-$\dot{m}$ models, where the full absorption photosphere extends to substantially larger radii.

The horizon total efficiency $\eta_H$ is also evaluated, which is defined as
\begin{equation}
    \eta_H = \frac{\dot{E}(r_H)+\dot{M}_{\rm acc} c^2}{\langle \dot{M}_{\rm acc}c^2\rangle_t}.
\end{equation}
When $\eta_H>100\%$,  net energy is extracted from the BH. The corresponding dimensionless spin-up parameter $s_H$ is defined as
\begin{equation}
    s_H \equiv \frac{{\rm d} j}{{\rm d} t} \frac{M_{\rm{}BH}}{\langle\dot{M}_{\rm acc} c^2 \rangle_t} = -j_H - 2a(1-\eta_H). 
\end{equation} \\[1ex]
Here, $j_H$ is the specific angular momentum flux of the accretion flow at the BH horizon, which is defined as
\begin{equation}
    j_H=\frac{\int T^r_{\phi}  (r_H) \,\rm dA_{\rm \theta\phi}}{\langle\dot{M}_{\rm acc} c^2 \rangle_t}
\end{equation}
Positive (negative) values of $s_H$ indicate the spin-up (spin-down) of the BH.

\begin{table*}
\centering
\caption{Additional model parameters. The wind luminosities are measured at $r=2,000\,r_g$ and are given in units of $L_{\rm Edd}$. We distinguish the radiative luminosity evaluated outside the effective photosphere for free-free absorption, $L_{\rm rad,ff}\equiv L_{\rm rad}(\tau_{\rm eff,ff}<1)$, from that evaluated outside the effective photosphere for full absorption, $L_{\rm rad,abs}\equiv L_{\rm rad}(\tau_{\rm eff,abs}<1)$. The corresponding radiative efficiencies are denoted by $\eta_{\rm rad,w}^{\rm ff}$ and $\eta_{\rm rad,w}^{\rm abs}$, respectively. The MRI quality factors $Q_{\theta,\rm MRI}$ and $Q_{\phi,\rm MRI}$ are evaluated at the radial grid cells at $r\simeq20\,r_g$ and $80\,r_g$ (see Appendix~\ref{ap:mri}).}
\label{tab:parameters}

\tabletypesize{\scriptsize}
\setlength{\tabcolsep}{3.0pt}

\begin{tabular}{lccccccccccccccccc}
\hline
\hline

\multicolumn{1}{c}{\multirow{2}{*}{Model}} &
\multirow{2}{*}{$a$} &
\multirow{2}{*}{$M_{\rm BH}$ [$M_\odot$]} &
\multirow{2}{*}{$\dot{M}_{\rm acc}$ [$\dot{M}_{\rm Edd}$]} &
\multicolumn{4}{c}{Wind Luminosity [$L_{\rm Edd}$]} &
\multicolumn{5}{c}{Efficiency [\%]} &
\multirow{2}{*}{$s_H$} &
\multicolumn{2}{c}{$r\simeq20\,r_g$} &
\multicolumn{2}{c}{$r\simeq80\,r_g$} \\ [0.8ex]
\cline{5-8}\cline{9-13}\cline{15-16}\cline{17-18}

& & & &
$L_{\rm kin}$ & $L_{\rm rad,ff}$ & $L_{\rm rad,abs}$ & $L_{\rm EM}$ &
$\eta_{\rm kin,w}$ & $\eta_{\rm rad,w}^{\rm ff}$ &
$\eta_{\rm rad,w}^{\rm abs}$ & $\eta_{\rm EM,w}$ & $\eta_H$ &
&
$Q_{\theta,\rm MRI}$ & $Q_{\phi,\rm MRI}$ &
$Q_{\theta,\rm MRI}$ & $Q_{\phi,\rm MRI}$ \\ [0.8ex]

\hline

\texttt{m5a0r1} & 0.0 & 5 & 21.8 & 7.1 & 3.4 & 3.4 & 1.3 & 3.3 & 1.5 & 1.5 & 0.6 & 6.2 & 1.2 & 43.4 & 22.1 & 47.0 & 13.2 \\
\texttt{m5a0r2} & 0.0 & 5 & 45.7 & 14.2 & 5.5 & 5.5 & 2.2 & 3.1 & 1.2 & 1.2 & 0.5 & 6.1 & 0.9 & 49.1 & 24.3 & 43.0 & 17.0 \\
\texttt{m5a0r3} & 0.0 & 5 & 172 & 57.6 & 30.3 & 28.7 & 14.5 & 3.4 & 1.8 & 1.7 & 0.8 & 5.6 & 0.7 & 81.1 & 27.8 & 29.8 & 15.0 \\
\texttt{m5a0r4} & 0.0 & 5 & 488 & 109 & 47.4 & 21.0 & 27.1 & 2.2 & 1.0 & 0.43 & 0.6 & 5.0 & 0.7 & 51.8 & 25.1 & 37.0 & 20.8 \\

\texttt{m5a9r1} & 0.9 & 5 & 5.4 & 11.7 & 4.5 & 4.5 & 6.9 & 21.7 & 8.3 & 8.3 & 12.7 & 78.5 & -4.8 & 45.2 & 14.2 & 11.0 & 7.4 \\
\texttt{m5a9r2} & 0.9 & 5 & 16.6 & 46.4 & 21.4 & 21.4 & 19.5 & 27.9 & 12.9 & 12.9 & 11.8 & 67.4 & -4.6 & 69.5 & 21.7 & 32.1 & 14.2 \\
\texttt{m5a9r3} & 0.9 & 5 & 120 & 696 & 181 & 180 & 311 & 57.8 & 15.0 & 14.9 & 25.8 & 137 & -7.5 & 49.5 & 21.9 & 40.9 & 16.9 \\
\texttt{m5a9r4} & 0.9 & 5 & 414 & 3026 & 775 & 688 & 1637 & 73.1 & 18.7 & 16.6 & 39.6 & 146 & -8.1 & 63.6 & 21.1 & 16.1 & 18.2 \\

\texttt{m15a0r1} & 0.0 & 15 & 7.7 & 3.1 & 2.1 & 2.1 & 0.73 & 4.1 & 2.8 & 2.8 & 1.0 & 8.6 & 1.7 & 46.3 & 14.5 & 10.7 & 6.7 \\
\texttt{m15a0r2} & 0.0 & 15 & 15.9 & 5.3 & 2.7 & 2.7 & 1.1 & 3.3 & 1.7 & 1.7 & 0.7 & 6.0 & 1.4 & 61.8 & 23.5 & 38.2 & 12.0 \\
\texttt{m15a0r3} & 0.0 & 15 & 65.1 & 25.8 & 9.6 & 9.6 & 3.9 & 4.0 & 1.5 & 1.5 & 0.6 & 6.5 & 0.9 & 42.2 & 23.1 & 46.4 & 17.0 \\
\texttt{m15a0r4} & 0.0 & 15 & 144 & 29.2 & 14.3 & 14.2 & 7.1 & 2.0 & 1.0 & 0.98 & 0.5 & 5.8 & 0.6 & 69.5 & 28.8 & 25.3 & 16.9 \\
\texttt{m15a0r5} & 0.0 & 15 & 195 & 51.3 & 28.5 & 25.4 & 13.1 & 2.6 & 1.5 & 1.3 & 0.7 & 6.3 & 0.7 & 53.4 & 23.3 & 42.2 & 21.9 \\
\texttt{m15a0r6} & 0.0 & 15 & 1438 & 364 & 224 & 76.6 & 92.9 & 2.5 & 1.6 & 0.53 & 0.6 & 6.3 & 0.7 & 46.9 & 25.1 & 39.6 & 24.9 \\
\texttt{m15a0r7} & 0.0 & 15 & 1592 & 539 & 249 & 35.1 & 135 & 3.4 & 1.6 & 0.22 & 0.9 & 5.9 & 0.6 & 74.7 & 26.1 & 27.3 & 16.4 \\

\texttt{m15a9r1} & 0.9 & 15 & 6.8 & 12.3 & 5.8 & 5.8 & 6.4 & 18.1 & 8.5 & 8.5 & 9.4 & 69.2 & -4.4 & 55.8 & 15.5 & 13.3 & 8.1 \\
\texttt{m15a9r2} & 0.9 & 15 & 10.2 & 27.9 & 11.6 & 11.6 & 13.4 & 27.4 & 11.4 & 11.4 & 13.2 & 75.7 & -4.9 & 53.0 & 20.1 & 33.4 & 12.2 \\
\texttt{m15a9r3} & 0.9 & 15 & 21.4 & 78.5 & 28.3 & 28.3 & 31.7 & 36.7 & 13.2 & 13.2 & 14.8 & 107 & -6.7 & 40.5 & 20.1 & 54.8 & 17.9 \\
\texttt{m15a9r4} & 0.9 & 15 & 37.0 & 131 & 57.2 & 57.1 & 53.3 & 35.3 & 15.4 & 15.4 & 14.4 & 107 & -7.7 & 31.8 & 19.2 & 58.9 & 18.4 \\
\texttt{m15a9r5} & 0.9 & 15 & 70.9 & 293 & 96.6 & 96.3 & 131 & 41.3 & 13.6 & 13.6 & 18.5 & 135 & -7.0 & 28.5 & 21.2 & 58.9 & 19.4 \\
\texttt{m15a9r6} & 0.9 & 15 & 191 & 1042 & 368 & 360 & 541 & 54.5 & 19.2 & 18.8 & 28.3 & 157 & -8.8 & 56.2 & 20.1 & 20.5 & 17.0 \\
\texttt{m15a9r7} & 0.9 & 15 & 450 & 2839 & 757 & 654 & 1313 & 63.0 & 16.8 & 14.5 & 29.1 & 126 & -6.6 & 63.7 & 26.2 & 36.2 & 15.0 \\
\texttt{m15a9r8} & 0.9 & 15 & 1168 & 8900 & 1883 & 1092 & 5306 & 76.2 & 16.1 & 9.4 & 45.4 & 141 & -8.7 & 56.2 & 19.3 & 11.1 & 18.2 \\

\texttt{m30a0r1} & 0.0 & 30 & 14.0 & 4.7 & 2.7 & 2.7 & 1.0 & 3.4 & 1.9 & 1.9 & 0.7 & 7.4 & 1.1 & 56.5 & 20.9 & 27.1 & 11.8 \\
\texttt{m30a0r2} & 0.0 & 30 & 158 & 43.2 & 16.4 & 15.9 & 6.5 & 2.7 & 1.0 & 1.0 & 0.41 & 5.6 & 0.7 & 82.2 & 26.3 & 65.7 & 18.7 \\
\texttt{m30a0r3} & 0.0 & 30 & 184 & 53.3 & 25.9 & 24.8 & 11.5 & 2.9 & 1.4 & 1.4 & 0.6 & 5.1 & 0.7 & 80.2 & 27.9 & 31.2 & 17.7 \\
\texttt{m30a0r4} & 0.0 & 30 & 1155 & 368 & 218 & 80.2 & 85.2 & 3.2 & 1.9 & 0.69 & 0.7 & 5.3 & 0.6 & 76.2 & 28.5 & 45.2 & 21.3 \\

\texttt{m30a9r1} & 0.9 & 30 & 10.2 & 36.5 & 10.6 & 10.6 & 18.1 & 35.6 & 10.4 & 10.4 & 17.6 & 94.1 & -6.1 & 39.9 & 18.4 & 21.5 & 12.8 \\
\texttt{m30a9r2} & 0.9 & 30 & 81.0 & 362 & 113 & 112 & 164 & 44.6 & 13.9 & 13.8 & 20.2 & 118 & -7.1 & 48.3 & 21.4 & 42.1 & 16.6 \\
\texttt{m30a9r3} & 0.9 & 30 & 130 & 762 & 223 & 220 & 394 & 58.8 & 17.2 & 17.0 & 30.4 & 150 & -8.4 & 31.8 & 20.8 & 37.5 & 18.0 \\

\hline
\end{tabular}
\end{table*}

\begin{table*}
\centering
\caption{Jet properties of the spinning ($a=0.9$) models at $r=2,000\,r_g$: jet efficiency $\eta_{\rm jet}$, total jet power $P_{\rm jet}$, and radiative, kinetic, and electromagnetic jet luminosities, $L_{\rm rad,jet}$, $L_{\rm kin,jet}$, and $L_{\rm EM,jet}$.}
\label{tab:jet_parameters}
\tabletypesize{\scriptsize}
\setlength{\tabcolsep}{3.0pt}

\begin{tabular}{lccccr}
\hline
\hline
Model & $\eta_{\rm jet}$ [\%] & $P_{\rm jet}$ [$L_{\rm Edd}$] & $L_{\rm rad,jet}$ [$L_{\rm Edd}$] & $L_{\rm kin,jet}$ [$L_{\rm Edd}$] & $L_{\rm EM,jet}$ [$L_{\rm Edd}$] \\ [1ex]
\hline

    \tt{m5a9r1} & 13.0 & 7.0 & 0.14 & 1.7 & 5.2 \\
	\tt{m5a9r2} & 12.1 & 20.1 & 0.70 & 4.5 & 14.9 \\
	\tt{m5a9r3} & 28.8 & 347 & 17.1 & 86.0 & 244 \\
	\tt{m5a9r4} & 20.4 & 843 & 54.4 & 253 & 535 \\
    
    \tt{m15a9r1} & 15.2 & 10.3 & 0.26 & 2.2 & 7.8 \\
    \tt{m15a9r2} & 19.0 & 19.3 & 0.54 & 4.7 & 14.1 \\
    \tt{m15a9r3} & 22.3 & 47.7 & 1.7 & 10.8 & 35.2 \\
    \tt{m15a9r4} & 29.3 & 108 & 6.8 & 25.0 & 76.7 \\
	\tt{m15a9r5} & 37.4 & 265 & 15.0 & 66.0 & 184 \\
	\tt{m15a9r6} & 34.3 & 656 & 36.1 & 149 & 471 \\
    \tt{m15a9r7} & 20.0 & 903 & 81.6 & 247 & 575 \\
    \tt{m15a9r8} & 12.2 & 1428 & 94.0 & 483 & 851 \\

	\tt{m30a9r1} & 21.2 & 21.7 & 0.53 & 5.3 & 15.9 \\
	\tt{m30a9r2} & 24.3 & 197 & 8.5 & 44.4 & 144 \\
	\tt{m30a9r3} & 36.2 & 469 & 37.5 & 116 & 316 \\
    
	\hline
\end{tabular}

\end{table*}


\nocite{Jiang.2019,Sadowski.2016,Yoshioka.2024,Kitaki.2021,Ohsuga.2009,Narayan.2017,Curd.2019,Curd.2023a}

\bibliography{export_bibtex}{}
\bibliographystyle{aasjournalv7}



\end{document}